\pdfoutput=1
\newcommand*{\ATLASLATEXPATH}{}
\documentclass[cernpreprint,USenglish,texlive=2011,txfonts]{\ATLASLATEXPATH atlasdoc}

\usepackage{\ATLASLATEXPATH atlaspackage}
\usepackage{\ATLASLATEXPATH atlasbiblatex}

\usepackage{\ATLASLATEXPATH atlascontribute}

\usepackage{\ATLASLATEXPATH atlasphysics}

\usepackage{multirow}
\usepackage{rotating}
\usepackage{tabularx}
\usepackage{epstopdf} 
\usepackage{mono_paper-defs}

\AtlasTitle{Search for new phenomena in final states with an energetic jet and large missing transverse momentum in $pp$
 collisions at $\sqrt{s}=13$~\TeV using the ATLAS detector}

\author{The ATLAS Collaboration}

\AtlasRefCode{EXOT-2015-03}

\PreprintIdNumber{CERN-EP-2016-075}

\AtlasJournal{Phys.\ Rev.\ D.}

\AtlasAbstract{%
 Results of a search for new phenomena in
  final states with an energetic jet and large missing transverse momentum are reported. 
The search 
 uses proton--proton collision data corresponding to an integrated luminosity of $3.2\,\ifb$ at $\sqrt{s}=13\TeV$ collected in 
  2015 with the ATLAS detector at the Large Hadron Collider. 
Events are required to have 
 at least one jet with a transverse momentum above $250\GeV$ and  no leptons.  Several signal regions 
are considered with increasing missing-transverse-momentum  
requirements between $\met >250\GeV$ and 
$\met > 700\GeV$. 
Good agreement is observed between the number of events in data and Standard Model predictions.
The results are translated into exclusion limits 
in models with large extra spatial dimensions,   
pair production of weakly interacting dark-matter 
candidates, and the production of supersymmetric particles in several 
compressed scenarios.
\\
\\
}

\hypersetup{pdftitle={ATLAS document},pdfauthor={The ATLAS Collaboration}}

\begin{document}

\maketitle

\tableofcontents

\section{Introduction}
\label{sec:intro}

Events with an energetic jet and large missing transverse momentum $\ptmi$ (with magnitude $\met$) in
the final state constitute a clean and distinctive signature in
searches for new physics beyond the Standard Model (SM) at colliders.  
Such signatures are referred to as monojet-like in this paper.
In particular, monojet (as well as   
monophoton and  mono-$W/Z$) final states have been
studied at the Large Hadron Collider (LHC)~\cite{Chatrchyan:2011nd,Chatrchyan:2012me,Chatrchyan:2012tea,Aad:2011xw,ATLAS:2012ky,Aad:2012fw,Khachatryan:2014rra,Khachatryan:2014rwa,Affolder:2000ef,Aad:2014nra,Aad:2014vea,Aad:2014vka,Aad:2013oja,Khachatryan:2014tva,Khachatryan:2014rra,Khachatryan:2015wza}
in the context of searches for large extra
spatial dimensions (LED),
supersymmetry (SUSY), 
and weakly interacting massive
particles (WIMPs) as candidates for dark matter. 

%
%

The Arkani-Hamed, Dimopoulos, and Dvali (ADD) model for LED~\cite{ArkaniHamed:1998rs}
explains the large difference between the electroweak unification scale at $O(10^2)\GeV$ and
the Planck scale   $M_{\rm{Pl}} \sim O(10^{19})\GeV$
by postulating the presence of $n$ extra spatial dimensions of size $R$,
and defining a fundamental Planck scale  in $4+n$ dimensions, $M_D$, given by
${M_{\rm{Pl}}}^2 \sim {M_D}^{2+n}R^n$. An appropriate
choice of $R$ for a given $n$ yields a value of $M_D$ at
the electroweak scale. The extra spatial dimensions are compactified,
 resulting in a Kaluza--Klein tower of massive graviton modes.
If produced in high-energy collisions in association with an energetic jet,
these graviton modes escape detection leading to a monojet-like signature in the final state.

%
%

Supersymmetry~\cite{Miyazawa:1966,Ramond:1971gb,Golfand:1971iw,Neveu:1971rx,Neveu:1971iv,Gervais:1971ji,Volkov:1973ix,Wess:1973kz,Wess:1974tw}
is a theory 
for physics beyond the SM that  
naturally solves the hierarchy
problem and provides a possible candidate for dark matter 
in the universe. SUSY enlarges the SM spectrum
of particles by introducing a new supersymmetric partner
(sparticle) for each particle in the SM. In particular,
a new scalar field is associated with each left- or right-handed
quark state and, ignoring inter-generational mixing, two squark mass eigenstates $\tilde{q}_1$
and $\tilde{q}_2$ result from the mixing of the scalar fields for a particular flavor.

In some
SUSY scenarios, 
a significant mass difference between the two eigenstates
in the bottom squark (sbottom) and top squark (stop) sectors 
can occur, leading to 
rather light sbottom $\tilde{b}_1$ and stop $\tilde{t}_1$ mass states. 
In addition, naturalness arguments suggest that the third generation squarks 
should be light, with masses below about $1\TeV$~\cite{Barbieri:1987fn}. 
In a generic 
supersymmetric extension of the SM   that assumes 
R-parity conservation~\cite{Fayet:1976et,Fayet:1977yc,Farrar:1978xj,Fayet:1979sa,Dimopoulos:1981zb}, sparticles are produced
in pairs and the lightest supersymmetric particle (LSP)
is stable. In this paper the LSP is assumed to be the lightest neutralino\footnote{
Neutralinos $\neuj$ ($j=1,2,3,4$ in the order of increasing mass) and charginos $\chaj$ ($j=1,2$)
are SUSY mass eigenstates formed from the mixing of the SUSY partners to the  Higgs
and electroweak gauge bosons.} $\ninoone$.  

\begin{figure}
\centering
\includegraphics[scale=1]{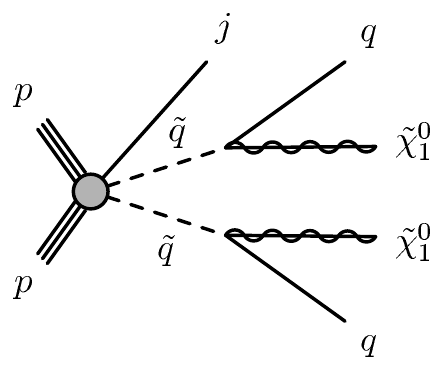}
\hspace{1cm}
\includegraphics[scale=1]{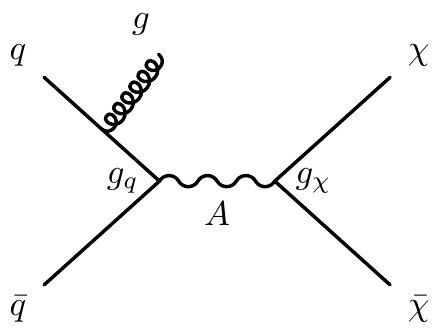}
\caption{Left: A generic diagram for the pair production of squarks with the decay mode $\tilde{q} \to q  + \ninoone$. 
Right: Diagram for the pair production of weakly interacting massive particles, with a leptophobic $Z'$-like mediator $A$ with axial-vector couplings exchanged in the $s$-channel. The presence of a jet from initial-state radiation
is indicated for both processes for illustration purposes.
}
\label{fig:feynman}
\end{figure}

The results from the monojet-like analysis are  
interpreted in terms of searches for squark production 
using simplified models 
in compressed scenarios for which
the mass difference
$\Delta m \equiv m_{\tilde{q}}                                                                                        
- m_{\ninoone}$ is small. Three separate processes are considered: stop pair production, where the stop
decays to a
charm quark and the LSP ($\tilde{t}_1 \to c + \ninoone$); sbottom pair production with $\tilde{b}_1 \to b  + \ninoone$; 
and squark  pair production, with $\tilde{q} \to q  + \ninoone$  $(q=u,d,c,s)$. For  relatively small 
 $\Delta m$,  both the transverse momenta of the quark
jets and the \met\ in the final state are low, making it  difficult to extract the signal from the large
multijet background.  In this study, the event selection makes use of the presence of
initial-state radiation (ISR) jets to identify signal events (see Fig.~\ref{fig:feynman} (left)).
In this case, the squark-pair system is boosted, leading to larger $\met$.

%
%

A nonbaryonic dark matter component in the universe is commonly used to
explain a range of astrophysical measurements~(see, for example,
Ref.~\cite{Bertone:2004pz} for a review).  Since none of the 
SM particles are adequate dark matter candidates, the existence of a new
particle is often hypothesized.
Weakly interacting massive particles 
are one such class of particle
candidates~\cite{Steigman:1984ac} that can be searched for at the
LHC. 
Such a new
particle would result in the correct relic density values for
nonrelativistic matter in the early universe~\cite{Kolb:1990vq}, as
measured by the Planck~\cite{Adam:2015rua} and WMAP~\cite{2013ApJS..208...19H}
satellites, if its mass is between a few \GeV{} and one \TeV{}
and if it has electroweak-scale interaction cross sections. Many new
particle-physics models such as SUSY~\cite{Miyazawa:1966,Ramond:1971gb,Golfand:1971iw,Neveu:1971rx,Neveu:1971iv,Gervais:1971ji,Volkov:1973ix,Wess:1973kz,Wess:1974tw}  
also predict WIMPs.

In contrast to the Run-1 analyses with the monojet-like final state~\cite{Aad:2015zva}, 
the results of this analysis are not interpreted in terms of the effective-field-theory models~\cite{Goodman:2010ku}. Simplified models are used instead, providing a more complete framework that involves new mediator particles between the SM and the Dark Sector~\cite{Abdallah:2015ter,LHCDMF:2015,Buchmueller:2014yoa,Boveia:2016mrp}.
The predictions from simplified models coincide with those obtained by using an 
effective-field-theory approach when the mediator mass considered is above $10\TeV$~\cite{Busoni:2013lha}. 
Here a model with an $s$-channel exchange of a spin-1 mediator particle with axial-vector couplings is considered, connecting the quarks to WIMPs of a Dirac fermion type. This is referred to as a leptophobic $Z'$-like model, and is defined by four free parameters:
the WIMP mass $m_\chi$, the mediator mass $m_A$, the coupling of the mediator to WIMPs ($g_\chi$) and the flavor-universal coupling to  quarks ($g_q$). Couplings to other SM particles are not allowed and the miminal mediator width is taken, defined in accord with Ref.~\cite{Buchmueller:2014yoa} as 

\begin{equation}
\Gamma_{\textrm{min}}=\frac{g_\chi^2 m_A}{12\pi} \beta_\chi^3 \theta(m_A-2m_\chi)
+ \sum_q \frac{3 g_q^2 m_A}{12\pi}\beta_q^3 \theta(m_A-2m_q)\;,
\end{equation}

where $\theta(x)$ denotes the Heaviside step function and $\beta_f=\sqrt{1-\frac{4m_f^2}{m_A^2}}$
is the velocity of the fermion $f$ with mass $m_f$ in the mediator rest frame. The sum runs over all 
quark flavors.
The monojet-like signature in this model emerges from initial-state radiation of a gluon as shown in Fig.~\ref{fig:feynman} (right).

%

The paper is organized as follows. The ATLAS detector is described in the next section. 
Section~\ref{sec:mc} provides details
of the simulations used in the analysis for background and signal processes.
Section~\ref{sec:recons} discusses the  reconstruction of jets, leptons, and  missing transverse momentum, while   
Section~\ref{sec:evt} 
describes the event selection.
The estimation of background contributions  and  
the study of systematic uncertainties are discussed in Sections~\ref{sec:backg} and~\ref{sec:syst}. 
The results are presented in  Section~\ref{sec:results} and are interpreted in terms 
of limits in models  for ADD LED,  SUSY in compressed scenarios,  and WIMP pair production.
Finally, Section~\ref{sec:sum} is devoted to the conclusions.

\section{Experimental setup}
\label{sec:atlas}

The ATLAS detector~\cite{Aad:2008zzm} covers almost the whole solid angle\footnote{The ATLAS experiment uses a right-handed
  coordinate system with its origin at the nominal interaction point
  (IP) in the center of the detector and the $z$-axis along the beam
  pipe. The $x$-axis points from the IP to the center of the LHC ring,
  and the $y$-axis points upward.  The azimuthal angle $\phi$ is
measured around the beam axis, and the polar angle $\theta$ is measured with respect to the $z$-axis.
The  transverse energy is defined as 
$E_{\rm T} = E \ {\rm sin}  \theta$, the 
transverse momentum as $\pt = p \ {\rm sin} \theta$,
and the pseudorapidity as $\eta = - {\rm ln}[{\rm tan}(\theta/2)]$.
The rapidity is defined as $y = 0.5 \times {\rm ln}[(E+p_z)/(E-p_z)]$, where $E$ denotes the energy and $p_z$ is 
the component of the momentum along the  beam direction.
}
around the collision point with layers of tracking detectors, calorimeters, and muon
chambers. The ATLAS inner detector  
 covers the pseudorapidity range $|\eta|<2.5$. 
It consists of a silicon pixel detector, a silicon microstrip detector, and a straw tube tracker that also measures transition radiation for particle identification, all immersed in a 2 T axial  magnetic field produced by a solenoid.
During the first LHC long shutdown, a new tracking layer, known as the Insertable B-Layer~\cite{Capeans:1291633}, 
was added at a radius of 33 mm.

High-granularity lead/liquid-argon (LAr) electromagnetic sampling calorimeters cover the pseudorapidity
range $|\eta|<$~3.2. The hadronic calorimetry in the range $|\eta|<$~1.7 is provided by a steel/scintillator-tile calorimeter, consisting of a large barrel and two smaller extended barrel cylinders, one on either side of
the central barrel. In the endcaps ($|\eta|>$~1.5), copper/LAr and tungsten/LAr hadronic
calorimeters match the outer $|\eta|$ limits of the endcap electromagnetic calorimeters. The LAr
forward calorimeters provide both the electromagnetic and hadronic energy measurements, and extend
the coverage to $|\eta| < 4.9$.

 The muon spectrometer measures the deflection of muons 
in the magnetic field provided by
large superconducting air-core toroid magnets
in the pseudorapidity range  $|\eta|<2.7$, instrumented with separate trigger and high-precision tracking chambers. 
Over most of the $\eta$ range, a measurement of the track coordinates in the bending direction of the magnetic 
field is provided
 by monitored drift tubes. Cathode strip chambers  with higher granularity are used in the innermost plane over $2.0 < |\eta| < 2.7$.
The muon    
fast trigger detectors cover the pseudorapidity range $|\eta| < 2.4$ and provide a measurement of the coordinate in the non-bending plane.

The data were collected using an online  two-level trigger system~\cite{ATL-DAQ-PUB-2016-001}  that selects events  of interest and reduces 
the event rate from several MHz to about 1\,kHz for recording and offline processing. 

\section{Monte Carlo simulation}
\label{sec:mc}

Monte Carlo (MC) simulated event samples are used to 
compute detector acceptance and reconstruction efficiencies,
determine signal and background contributions, and estimate systematic uncertainties in the final results. 
Background contributions from multijet processes are determined directly from data. 

\subsection{Background simulation}


The expected background to the monojet-like signature is dominated by $\znn$+jets and   
$W$+jets production with $\wtn$+jets  being the largest  $W$+jets background, and  includes small contributions from $\zll$+jets ($\ell = e, \mu, \tau$), multijet, $t\bar{t}$, single-top, 
and diboson ($WW,WZ,ZZ$) processes. Contributions from 
top production  associated with  additional vector
bosons ($\ttb + W$, $\ttb + Z$, or $t + Z + q/b$ processes) are negligible.


Events containing $W$ or $Z$ bosons with associated jets are simulated using the {SHERPA}-2.1.1~\cite{sherpa} generator. Matrix elements (ME) are calculated for up to two partons at next-to-leading order (NLO) and four partons at leading order  (LO) using the Comix~\cite{Gleisberg:2008fv} and OpenLoops~\cite{Cascioli:2011va} matrix element generators and merged with the {SHERPA} parton shower (PS)~\cite{Schumann:2007mg} using the ME+PS@NLO prescription~\cite{Hoeche:2012yf}. The CT10~\cite{ct10} parton distribution function (PDF) set is used in conjunction with a dedicated parton shower tuning developed by the authors of SHERPA. 
The MC predictions are initially
normalized to next-to-next-to-leading-order (NNLO) perturbative QCD (pQCD) predictions according to
DYNNLO~\cite{Catani:2009sm,Catani:2007vq}
using  MSTW2008 90$\%$ CL NNLO PDF sets~\cite{mstw}.

For the generation of $\ttb$ and single top-quarks in the $Wt$-channel and $s$-channel the POWHEG-BOX v2~\cite{powheg} generator with the CT10 PDF sets in the matrix element calculations is used. Electroweak $t$-channel single top-quark events are generated using the POWHEG-BOX v1 generator. This generator uses the 
four-flavor scheme 
for the calculations of NLO matrix elements with the fixed
four-flavor PDF set CT10. The parton shower, fragmentation, and underlying event are simulated using PYTHIA-6.428~\cite{pythia} with the CTEQ6L1~\cite{Pumplin:2002vw}  PDF sets and the corresponding 
Perugia 2012 set of tuned parameters (P2012 tune)~\cite{Skands:2010ak}. The top-quark mass is set to $172.5\GeV$. The EvtGen v1.2.0 program~\cite{EvtGen} is used 
to model the decays of the bottom and charm hadrons.
Finally, diboson samples ($WW$, $WZ$, and $ZZ$ production) are generated using
 {SHERPA}-2.1.1 with CT10 PDFs and are normalized 
to NLO pQCD predictions~\cite{Campbell:2011bn}.  The diboson samples 
are also generated using {POWHEG} interfaced to  {PYTHIA}-8.186 and using CT10 PDFs  
for studies of systematic uncertainties.


\subsection{Signal simulation}

Simulated samples for the ADD LED model with 
different numbers of extra dimensions in the range $n = 2$--$6$ and $M_D$ in the range $2$--$5$~TeV 
are generated using {PYTHIA}-8.165  with
NNPDF23LO~\cite{Ball:2012cx}  PDFs. The renormalization scale is set to
the geometric mean of the transverse mass of the two produced particles,  
$\sqrt{(p_{\rm{T},G}^2 + m_G^2)(p_{\rm{T},p}^2 + m_p^2)}$, 
 where $m_G$ and $p_{\rm{T},G}$ ($m_p$ and $p_{\rm{T},p}$) denote, respectively, the mass and the  
transverse momentum of the graviton (parton)  in the final state. The factorization scale is
set to the minimum transverse mass $\sqrt{m^2 + \pt^2}$ of the graviton and the parton. 
 
SUSY signals  for
stop pair production with $\tilde{t}_1 \to c + \tilde{\chi}_{1}^{0}$, for  
sbottom pair production decaying
as $\tilde{b}_1 \to b + \ninoone$, and 
for
the production of squark pairs from the first two squark generations with
$\tilde{q} \to q + \ninoone$ ($q =u,d,c,s$) are considered. 
Events are generated with MG5\_aMC@NLO v5.2.2.3~\cite{Alwall:2014hca} interfaced to {PYTHIA}-8.186
with the ATLAS A14~\cite{tune14} tune 
for the modeling of the squark decay, and the parton showering,
hadronization, and underlying event. The matrix element  calculation is performed at tree level,
and includes the emission of up to two additional partons. The renormalization and factorization scales
are set to the sum of transverse masses of all final state particles.
The PDF used for the
generation is NNPDF23LO. The ME–PS matching is done using the CKKW-L~\cite{Lonnblad:2011xx} prescription,
with a matching scale set to one quarter of the pair-produced superpartner mass. 
Simulated samples with squark masses in the range between $250\GeV$ and $700\GeV$ and $\Delta m$ varying between $5\GeV$ and 
$25\GeV$ are produced.  
Signal cross sections are calculated to NLO 
in the strong coupling constant, adding the resummation of soft gluon
emission at next-to-leading-logarithmic 
(NLO+NLL) accuracy~\cite{Beenakker:1997ut,Beenakker:2010nq,Beenakker:2011fu}. The
nominal cross section and its uncertainty are taken from an envelope
of cross-section predictions using different PDF sets and
factorization and renormalization scales, as described in
Ref.~\cite{Borschensky:2014cia}.

WIMP signals are simulated in POWHEG-BOX v2~\cite{Alioli:2010xd,Frixione:2007vw,Nason:2004rx} using revision 3049 of the DMV model 
implementation of WIMP pair production with $s$-channel spin-1 mediator exchange
at NLO precision including parton showering effects, introduced in Ref.~\cite{Haisch:2013ata}.
Renormalization and factorization scales are set to $H_\textrm{T}/2$ on an event-by-event basis, where $H_\textrm{T}=\sqrt{m_{\chi\chi}^2+p_{\textrm{T}, j1}^2}+p_{\textrm{T}, j1}$ is defined by the invariant mass of the WIMP pair ($m_{\chi\chi}$) and the transverse momentum of the hardest jet ($p_{\textrm{T}, j1}$).
A Breit--Wigner distribution is chosen to describe the mediator propagator. Events are generated using the NNPDF30NLO~\cite{Ball:2014uwa} parton distribution functions and interfaced to {PYTHIA}-8.205 with the ATLAS A14 tune for parton showering.
Couplings of the mediator to WIMPs and  quarks are set to $g_\chi=1$ and $g_q=1/4$, leading to narrow mediators with $\Gamma_{\textrm{min}}/m_A$ up to about $5\%$. A grid of samples is produced for WIMP masses ranging from $1\GeV$ to $1\TeV$ and mediator masses between $10\GeV$ and $2\TeV$.\footnote{In the generation of the samples, the {\it bornktmin} and {\it bornsuppfact} MC parameters~\cite{Alioli:2010xd} are set to $150\GeV$ and $1\TeV$, respectively, in order to suppress the generation of events at low \MET{}.}

%
%
Differing pileup (multiple proton--proton interactions in the same or neighboring bunch-crossings) 
conditions as a function of the instantaneous luminosity are taken into account by overlaying simulated 
minimum-bias events generated with   {PYTHIA} onto the hard-scattering process. 
The MC-generated samples are processed  with a full ATLAS detector simulation~\cite{Aad:2010ah} based on
the GEANT4 program~\cite{Agostinelli:2002hh}.   
The simulated events are
reconstructed and analyzed with the same analysis chain as for the data, using the same trigger and event 
selection criteria.

\section{Reconstruction of physics objects}
\label{sec:recons}

Jets are reconstructed from energy deposits in the calorimeters using 
the $\akt$ jet algorithm~\cite{paper:antikt} with the radius parameter (in $y$--$\phi$ space) 
set to 0.4. 
The measured 
jet transverse momentum  
is corrected for detector effects,  including the noncompensating character of the calorimeter, by weighting
 energy deposits arising from electromagnetic and hadronic showers differently.   
In addition, jets are corrected for 
contributions from pileup,  
as described in Ref.~\cite{Aad:2011he}.  Jets with  corrected $p_{\rm T} > 20\GeV$ and $|\eta| <2.8$ are initially considered
 in the analysis.
Track-based variables to suppress pileup jets have been developed.
A combination of two such variables called the jet-vertex tagger (JVT) is constructed. 
In order to remove jets originating from pileup collisions, for central jets ($|\eta| < 2.4$) with $p_{\rm T} < 50\GeV$ 
a significant fraction of the tracks associated with each jet must have an origin compatible with the primary vertex, as defined
by the jet-vertex tagger~\cite{ATLAS-CONF-2014-018}.

The presence of leptons (electrons or muons) in the final state  is used in the analysis 
to define control samples and to reject background contributions in the signal regions (see Sections~\ref{sec:evt} and~\ref{sec:backg}). 
Electron candidates are initially required to have
$p_{\rm T} > 20\GeV$ and $|\eta| <2.47$, and to satisfy the loose
 electron shower shape and track selection criteria described
in Refs.~\cite{Aad:2016jkr,ATLAS-CONF-2014-032}.
Overlaps between identified electrons and jets in the final state are
resolved. Jets are discarded if their separation $\Delta R = \sqrt{(\Delta \eta)^2 + (\Delta \phi)^2}$ from
an identified electron is less than~0.2.
The electrons separated by $\Delta R$
between 0.2 and  0.4 from any remaining jet are removed.

Muon candidates are formed by combining information from the muon spectrometer and inner 
tracking detectors as described in
Ref.~\cite{Aad:2016jkr} and are 
required to have  $p_{\rm T} > 10\GeV$ and  $|\eta| < 2.5$.
Jets with $p_{\rm T} > 20\GeV$ and less than  three tracks with $\pt > 0.4\GeV$ 
associated with them are discarded if their separation $\Delta R$ from
an identified muon is less than~0.4. 
The muon is discarded if it is matched to a jet that has at least three tracks associated  with it.

The $\met$ is reconstructed using all energy deposits in the calorimeter up 
to pseudorapidity $|\eta| =  4.9$. 
Clusters associated with either electrons or photons with $\pt >20\GeV$ and those
associated with jets with $\pt >20\GeV$ make use of the corresponding calibrations for these objects. Softer jets 
and clusters not associated with these objects are calibrated using tracking information~\cite{Aad:2012re}.   
As discussed below,  in this analysis the $\met$ is not corrected for the presence of muons in the final state.

\section{Event selection}
\label{sec:evt}

The data sample considered in this paper was collected with
tracking detectors, calorimeters, muon chambers, and magnets 
fully operational, and corresponds  
to a total integrated luminosity of $3.2\,\ifb$. 
The data were selected online using 
a trigger logic that selects events with  
$\met$ above $70\GeV$, as computed at the final stage of the two-level
trigger system of ATLAS.
With the final analysis requirements,  
the trigger selection is fully  efficient for $\met > 250\GeV$, 
as determined using a data sample with muons in the final state.
The following selection criteria,
summarized in Table~\ref{tab:sr}, are applied in the signal regions.

\begin{itemize}

\item Events are required to have a
reconstructed primary vertex for the interaction with at least two associated tracks with $\pt>0.4\GeV$ 
and consistent with the beamspot envelope;
when more than one such vertex is found, the vertex with the largest summed $\pt^2$
of the associated tracks is chosen.

\item Events are required to have $\met > 250\GeV$.
The analysis selects events with
a leading (highest $\pt$) jet  with $\pt > 250\GeV$ and $|\eta| < 2.4$ in the final state.   
A maximum of four  jets with $\pt > 30\GeV$ and $|\eta| < 2.8$ are allowed. 
A separation in the azimuthal plane of $\Delta\phi(\textrm{jet},{\ptmi}) > 0.4$ 
between the missing transverse momentum direction and each selected jet is required.
This requirement reduces the multijet background contribution where the
  large $\met$  originates  mainly from jet energy mismeasurement.

\item Events are rejected if they contain any jet inconsistent with the requirement that they originate from a proton--proton collision.
Jet quality selection criteria~\cite{clean} involve
quantities such as the 
pulse shape of the energy depositions in the cells of the calorimeters,  
electromagnetic fraction in the calorimeter, calorimeter sampling fraction, or charged-particle fraction.\footnote{
The charged-particle fraction is defined as
$f_{{\textrm{ch}}}=\sum p_{\rm T}^{{\textrm{track,jet}}} /p_{\rm T}^{{\textrm{jet}}}$,
where $\sum p_{\rm T}^{{\textrm{track,jet}}}$ is the scalar
sum of the transverse momenta of tracks associated with the primary vertex
within a cone of radius $\Delta R=0.4$ around the jet axis, and
$p_{\rm T}^{{ \textrm{jet}}}$ is the transverse momentum as determined
from calorimetric measurements. }
The loose criteria are applied to all jets with $\ptjet > 20\GeV$ and $|\eta| < 2.8$,
dealing efficiently with coherent noise and electronic noise bursts in the calorimeter producing anomalous energy depositions~\cite{Aad:2013zwa}.
Non-collision backgrounds, i.e. energy depositions in the calorimeters due to muons of beam-induced or cosmic-ray origin,
are further suppressed by applying the tight selection criteria to the leading jet: the ratio of 
the jet charged-particle fraction to the calorimeter sampling fraction,\footnote{
$f_{\small{\textrm{max}}}$ denotes                                                            
the maximum fraction of the jet energy collected by a single  calorimeter layer.}              
$f_{{\textrm{ch}}}/f_{\small{\textrm{max}}}$,  
is required to be larger than 0.1. 
These requirements have a negligible effect on the signal efficiency. 
 
\item Events with identified muons with $\pt > 10\GeV$ or electrons with $\pt > 20\GeV$ in the final state are vetoed.

\end{itemize}

Inclusive (IM1--IM7) and exclusive (EM1--EM6) signal regions are 
considered with increasing $\met$ thresholds from $250\GeV$ to $700\GeV$ (see Table~\ref{tab:sr}).
The use of inclusive $\met$ signal regions follows the Run 1 strategy, where the  results are translated into model-independent cross section upper limits for the production of new physics.  The use of exclusive $\met$ signal regions effectively explores information from the shape of the $\met$ distribution (see Sections~\ref{sec:fitback} and \ref{sec:results}) and enhances the sensitivity to the different new physics models.


\begin{table*}[!ht]
\renewcommand{\baselinestretch}{1}
\caption{Event selection criteria applied, as described in Section~\ref{sec:evt}.}
\begin{center}
\begin{footnotesize}
\begin{tabular*}{\textwidth}{@{\extracolsep{\fill}}lccccccc}\hline
\multicolumn{8}{c}{\small{Selection criteria}} \\\hline\hline
\multicolumn{8}{l}{Primary vertex}\\
\multicolumn{8}{l}{$\met > 250\GeV$ }\\
\multicolumn{8}{l}{Leading  jet with $\pt >250\GeV$ and $|\eta|< 2.4$}\\
\multicolumn{8}{l}{At most four jets with $\pt > 30\GeV$ and $|\eta|<2.8$}\\
\multicolumn{8}{l}{$\Delta\phi(\textrm{jet},\ptmi) > 0.4$}\\ 
\multicolumn{8}{l}{Jet quality requirements}   \\
\multicolumn{8}{l}{No identified muons with $\pt > 10\GeV$ or electrons with $\pt > 20\GeV$}\\ \hline
Inclusive signal region        & IM1 & IM2 & IM3 & IM4 & IM5 & IM6 & IM7 \\
$\met$ (\GeV) & $> 250$  & $> 300$  & $> 350$ & $> 400$ & $> 500$ & $> 600$  & $> 700$ \\ \hline
Exclusive signal region        & EM1 & EM2 & EM3 & EM4 & EM5 & EM6 & \\ 
$\met$ (\GeV) & [250--300] & [300--350]  & [350--400] & [400--500] & [500--600] & [600--700]  & \\  \hline
\end{tabular*}
\end{footnotesize}
\end{center}
\label{tab:sr}
\end{table*}

\section{Background estimation}
\label{sec:backg}

The $W$+jets, $\znn$+jets, $\ztt$+jets, and $\zmm$+jets  backgrounds are constrained using 
MC samples normalized with data in selected control regions.
The normalization factors are
extracted simultaneously using a global fit that includes systematic uncertainties,
to properly take into account correlations.

A $\wmn$+jets control sample is used to define normalization factors                                                      
for $\wmn$+jets and $\znn$+jets processes.                                              
As discussed in Section~\ref{sec:fitback}, 
the use of the $\wmn$+jets control sample to constrain the normalization of the                                            
$\znn$+jets process                                                                                                        
translates into a reduced uncertainty in the estimation                                                                    
of the main irreducible background contribution,  due to a partial cancellation of systematic uncertainties and the        
statistical power of the                                                                                                   
$\wmn$+jets control sample in data,  which is about seven times larger than the $\zmm$+jets control sample.                
A $\wen$+jets control sample is used to constrain the normalization of the 
$\wen$+jets and $\wtn$+jets background processes. For the latter, 
this is motivated by the fact that 
the $\tau$ lepton in the $\wtn$+jets background process mainly decays hadronically 
leading to a final-state topology in the detector similar to that of the $\wen$+jets sample. 
A small  $\ztt$+jets background contribution is also constrained using the $\wen$+jets control sample. 
Uncertainties related to the difference between $W$+jets and $Z$+jets final states, leading to potential differences in    
event kinematics and selection acceptances and efficiencies,  are discussed in Section~\ref{sec:syst}.                     
Finally, a $\zmm$+jets control sample is used to constrain the $\zmm$+jets background contribution.

The remaining SM backgrounds from $\zee$+jets,\footnote{
In the course of the analysis, the use of an additional $\zee$+jets control
sample was explored for constraining the $\zee$+jets and $\znn$+jets background contributions, 
leading to an insignificant improvement in the background determination.
}  $t\bar{t}$, single top, and dibosons 
are determined using MC simulated samples, while the multijet background  contribution
is extracted from data.
The  
contributions from non-collision backgrounds are
estimated in data using the beam-induced background identification techniques described in Ref.~\cite{Aad:2013zwa}.

The methodology and the samples used for estimating the background are summarized in Table~\ref{tab:CRs}. 
In the following subsections, details 
of the definition of the $W/Z$+jets control regions and of the data-driven determination of
the multijet and beam-induced backgrounds are given. This is followed by a description of the 
background fits. 

%
%

\begin{table}[!ht]
\caption{
Summary of the methods and control samples used to constrain the different background contributions in the signal regions. 
}
\begin{center}
\begin{footnotesize}
\begin{tabular*}{\textwidth}{@{\extracolsep{\fill}}lcc}\hline
{Background process}     & Method       &  {Control sample}  \\ \hline
$\znn$+jets    & MC and control samples in data  & $\wmn$  \\          
$\wen$+jets    & MC and control samples in data  & $\wen$  \\             
$\wtn$+jets    & MC and control samples in data  & $\wen$ \\            
$\wmn$+jets    & MC and control samples in data  & $\wmn$ \\           \hline 
$\zmm$+jets    & MC and control samples in data  & $\zmm$ \\
$\ztt$+jets    & MC and control samples in data  & $\wen$ \\
$\zee$+jets    &   MC only & \\
$\ttb$, single top &  MC only & \\           
Diboson            &  MC only & \\         
Multijets          &  data-driven & \\
Non-collision      &  data-driven & \\ \hline\hline
\end{tabular*}
\end{footnotesize}
\end{center}
\label{tab:CRs}
\end{table}


\subsection{$W/Z$+jets background}

Control samples in data,  
with identified electrons or muons in the final state  and with requirements on 
the jet $\ptjet$ and $\met$ identical to those in the signal regions,  are used to determine 
the  $\wln$+jets ($\ell = e,\mu,\tau$), $\znn$+jets, and $\zll$+jets ($\ell = \mu,\tau$)  background contributions. 
The $\zee$+jets           background contribution           is tiny    and it is determined from MC simulation. 
The $\met$-based online trigger used in the analysis 
does not include muon information in the $\met$ calculation. This allows the 
collection of $\wmn$+jets and $\zmm$+jets control samples with the same trigger as for the signal regions.


A $\wmn$+jets
control sample is selected by requiring a muon consistent with originating from the primary vertex with $\pt > 10\GeV$, and  
transverse mass
in the range  $30\GeV<  m_{\rm T} <  100\GeV$. 
The transverse mass $m_{\rm T} = \sqrt{2\pt^{\ell}\pt^{\nu}[1-\cos(\phi^{\ell}-\phi^{\nu})]}$  is defined by the lepton  and neutrino transverse momenta,   
where the $(x,y)$ components of the neutrino momentum are taken to be
the same as the corresponding $\ptmi$ components. Events with identified electrons in the final state are vetoed. 
Similarly, a $\zmm$+jets control
sample is selected by  requiring the presence of two muons with $\pt >10\GeV$ and 
invariant mass in the range $66\GeV < m_{\mu \mu} < 116\GeV$. In the $\wmn$+jets and  
$\zmm$+jets control regions,  the $\met$ is not corrected for the presence of 
the muons in the final state,  
motivated by the fact that these control regions
are used to estimate the $\znn$+jets and the $\zmm$+jets backgrounds, respectively, 
in the signal regions with no identified muons.         

Finally, a $\wen$+jets dominated control sample is defined 
with an isolated electron candidate with $\pt >20\GeV$, selected with tight or medium selection criteria~\cite{Aad:2016jkr,ATLAS-CONF-2014-032} depending on $\pt$, and 
no additional identified leptons in the final state.
The $\met$ calculation
includes the contribution of the energy cluster from the identified electron in the calorimeter (no attempt is made to subtract it), 
since $\wen$+jets processes contribute to the  background in the signal regions 
when the electron is not identified.

Monte Carlo-based scale factors, determined from the
{SHERPA} simulation, are defined for each of the signal
selections to estimate the different  background contributions in the signal regions.
As an illustration, in the case of the dominant $\znn$+jets background
process its contribution to a given signal region 
$N^{\znn}_{\rm{signal}}$ 
is determined using the  $\wmn$+jets control sample in data 
according to

\begin{equation}
N^{\znn}_{\rm{signal}} = (N^{\rm{data}}_{\wmn,{\rm{control}}} - N^{{\rm{non}}-W}_{\wmn,{\rm{control}}})  \times
\frac{N^{{\rm{MC}} (\znn)}_{\rm{signal}}}{N^{{\rm{MC}}}_{\wmn,{\rm{control}}}}, 
\end{equation}
\noindent

\noindent
where $N^{{\rm{MC}} (\znn)}_{\rm{signal}}$ denotes the background predicted 
by the MC simulation in the signal region, and $N^{\rm{data}}_{\wmn,{\rm{control}}}$,  
 $N^{\rm{MC}}_{\wmn,{\rm{control}}}$, and $N^{\rm{non-W}}_{\wmn,{\rm{control}}}$ denote, in the control region, the 
number of data events, the number of $\wmn$+jets  candidates from MC simulation,  and 
the  non-$\wmn$ background contribution, respectively. 
The $N^{\rm{non-W}}_{\wmn,{\rm{control}}}$ term  refers mainly to top-quark and diboson processes,  but also 
includes  contributions from other $W/Z$+jets processes. Multijets and non-collision backgrounds in 
the control regions are negligible.  

As discussed in Section~\ref{sec:fitback}, a global simultaneous likelihood 
fit to all the control regions
is used to determine the normalization factors.

\subsection{Multijets background}
\label{sec:qcdback}

The multijet background with large $\met$ mainly originates 
from the misreconstruction of the energy of a 
jet in the calorimeter and to a lesser extent is due to the presence 
of neutrinos in the final state from heavy-flavor hadron decays.  
In this analysis, the multijet background is determined from data, using the jet smearing method as
described in Ref.~\cite{Aad:2012fqa}, which relies  on the assumption that the $\met$ of multijet events is dominated by  
fluctuations in the jet response in the detector which can be measured in the data. 
For the IM1 and EM1 selections, the multijets background constitutes about $0.5\%$ of the total background, and  
 is negligible for the other signal regions.

\subsection{Non-collision background}
\label{sec:noncoll}

Non-collision backgrounds 
represent a significant portion of data acquired by $\MET$ triggers.
These backgrounds resemble the topology of monojet-like final states and a dedicated strategy with a suppression power of approximately $10^3$ is needed in order to reduce these backgrounds to a sub-percent level. This is achieved by the jet quality selection criteria described in Section~\ref{sec:evt}.
The rate of jets due to cosmic-ray muons surviving this selection, as measured in dedicated cosmic ray datasets, is found to be negligible compared to the rate of data in the monojet-like signal regions.
The main source of residual non-collision backgrounds is therefore beam-induced muons
originating in the particle cascades due to beam halo protons intercepting the LHC collimators.
The non-collision background is estimated using a method that identifies beam-induced muons based on the spatial matching of calorimeter clusters to muon track segments, reconstructed in the muon-system endcaps and pointing in a direction nearly parallel to the beam pipe~\cite{Aad:2013zwa}.
The number of events where the reconstructed objects satisfy the identification criteria is corrected for the efficiency of this method. The efficiency is evaluated in a dedicated beam-induced background-enhanced region defined by inverting the tight jet quality selection imposed on the leading jet.

The results indicate an almost negligible contribution from non-collision backgrounds in the signal regions. 
As an example, 110 and 19 non-collision background events are estimated in the IM1 and EM3 
signal regions, respectively, with no sign of non-collision backgrounds at $\met>500\gev$. This constitutes
about 0.5$\%$ of the total background for the IM1 and EM3 selections.

\subsection{Background fits}
\label{sec:fitback}

The use of control regions to constrain the normalization of the dominant background contributions from 
$\znn$+jets and  $W$+jets  significantly reduces
the relatively large  theoretical and experimental 
systematic uncertainties, of the order of  $20\%$--$40\%$,  associated with purely MC-based 
background predictions in the signal regions.
A complete study of systematic uncertainties is carried out, as detailed in Section~\ref{sec:syst}.
To determine the final uncertainty in the total background,  
all systematic uncertainties are treated as nuisance
parameters with Gaussian shapes in a fit based on the profile
likelihood method~\cite{statforumlimits} and which takes into account correlations among systematic variations.
The likelihood also takes into account cross-contamination between different background sources in the control regions. 

A simultaneous likelihood fit to the $\wmn$+jets, $\wen$+jets, and 
$\zmm$+jets control regions is performed to normalize and
constrain the corresponding background estimates in the signal regions.
Background-only fits are performed separately in each of the inclusive regions IM1--IM7, as described in Section~\ref{sec:evt}. 
In addition, a fit simultaneously using all the exclusive $\met$ regions EM1--EM6 and IM7 is performed. 
In this case, normalization factors are considered separately in each exclusive $\met$ region, which 
effectively employs information from the shape of the $\met$ distribution to 
enhance  the sensitivity of the analysis to the presence of new phenomena.

The results of the background-only fit in the control regions 
are presented in detail in Table~\ref{tab:fitm1} for the IM1 selection. 
Tables~\ref{tab:wen}--\ref{tab:zmm} collect the results for the total background predictions in each of the  control regions for  
the inclusive and exclusive $\met$ selections. As the tables indicate, the $W/Z$+jets background predictions 
receive multiplicative  normalization factors that vary in the range 
between 0.8 and 1.2, depending on the process and the kinematic selection. Good agreement is observed between the 
normalization factors obtained by using inclusive or exclusive $\met$ regions.

%
%

\begin{table}[tbh]
\caption{Data and background predictions in the control regions before and after the fit is performed for the IM1 selection.
The background predictions include both the statistical and systematic uncertainties.
The individual uncertainties are correlated, and do not necessarily add 
in quadrature to the total background uncertainty.
}
\begin{center}
\setlength{\tabcolsep}{0.0pc}
{\footnotesize
\begin{tabular*}{\textwidth}{@{\extracolsep{\fill}}lrrr}
\noalign{\smallskip}\hline\noalign{\smallskip}
{\bf  IM1 control regions}             & $\wen$             & $\wmn$            & $\zmm$             \\
\noalign{\smallskip}\hline\noalign{\smallskip}
Observed events (3.2~fb${}^{-1}$)      & $3559$             & $10481$           & $1488$                    \\
\noalign{\smallskip}\hline\noalign{\smallskip}
SM prediction (post-fit)               &  $3559 \pm 60$     &  $10480 \pm 100$  &  $1488 \pm 39$ \\ 
\noalign{\smallskip}\hline\noalign{\smallskip}
        Fitted $\wen$                  &  $2410 \pm 140$    &  $0.4 \pm 0.1$    &  $-$               \\ 
        Fitted $\wmn$                  &  $2.4 \pm 0.3$     &  $8550 \pm 330$   &  $1.8 \pm 0.3$ \\ 
        Fitted $\wtn$                  &  $462 \pm 27$      &  $435 \pm 28$     &  $0.14 \pm 0.02$ \\ 
        Fitted $\zee$                  &  $0.5 \pm 0.1$     &  $-$              &  $-$               \\ 
        Fitted $\zmm$                  &  $0.02 \pm 0.02$   &  $143 \pm 10$     &  $1395 \pm 41$ \\ 
        Fitted $\ztt$                  &  $30 \pm 2$        &  $22 \pm 4$       &  $0.5 \pm 0.1$ \\ 
        Fitted $\znn$                  &  $1.8 \pm 0.1$     &  $2.3 \pm 0.2$    &  $-$               \\ 
        Expected $\ttbar$, single top  &  $500 \pm 150$     &  $1060 \pm 330$   &  $42 \pm 13$ \\ 
        Expected dibosons              &  $150 \pm 13$      &  $260 \pm 25$     &  $48 \pm 5$ \\ 
\noalign{\smallskip}\hline\noalign{\smallskip}
MC exp. SM events                      &  $3990 \pm 320$    &  $10500 \pm 710$  &  $1520 \pm 98$ \\ 
\noalign{\smallskip}\hline\noalign{\smallskip}
        Fit input $\wen$               &  $2770 \pm 210$    &  $0.4 \pm 0.1$    &  $-$               \\ 
        Fit input $\wmn$               &  $2.4 \pm 0.3$     &  $8500 \pm 520$   &  $1.8 \pm 0.2$ \\ 
        Fit input $\wtn$               &  $531 \pm 39$      &  $500 \pm 34$     &  $0.16 \pm 0.03$ \\ 
        Fit input $\zee$               &  $0.5 \pm 0.1$     &  $-$              &  $-$               \\ 
        Fit input $\zmm$               &  $0.02 \pm 0.02$   &  $146 \pm 13$     &  $1427 \pm 92$ \\ 
        Fit input $\ztt$               &  $34 \pm 3$        &  $25 \pm 4$       &  $0.6 \pm 0.1$ \\ 
        Fit input $\znn$               &  $1.8 \pm 0.1$     &  $2.2 \pm 0.1$    &  $-$               \\ 
        Fit input $\ttbar$, single top &  $500 \pm 160$     &  $1060 \pm 340$   &  $42 \pm 13$ \\ 
        Fit input dibosons             &  $150 \pm 13$      &  $260 \pm 25$     &  $48 \pm 5$ \\ 
\noalign{\smallskip}\hline\noalign{\smallskip}
\end{tabular*}
}
\end{center}
\label{tab:fitm1}
\end{table}

\begin{table}[!ht]
\caption{Data and SM background prediction, before and after the fit, in the $\wen$ control region for the different selections.
For the SM predictions both the statistical and systematic uncertainties are included.
}
\begin{center}
\begin{footnotesize}
\begin{tabular*}{\textwidth}{@{\extracolsep{\fill}}lrrrrrrr}\hline
{\bf  Inclusive Selection}              & IM1 & IM2 & IM3 & IM4 & IM5 & IM6 & IM7 \\
Observed events  (3.2 fb${}^{-1}$)& $3559$         & $1866$         & $992$         & $532$        & $183$        & $72$       & $32$ \\ 
SM prediction  (post-fit)         & $3559 \pm 60$  &  $1866 \pm 43$ &  $992 \pm 32$ &  $532 \pm 23$ &  $183 \pm 14$ &  $72 \pm 8$ &  $32 \pm 6$  \\ 
SM prediction (pre-fit)           &  $3990 \pm 320$ &   $2110 \pm 170$ &   $1142 \pm 94$ &   $654 \pm 54$ &   $216 \pm 19$ &   $85 \pm 8$ &  $34 \pm 3$   \\ \hline
{\bf  Exclusive Selection}              & EM1 & EM2 & EM3 & EM4 & EM5 & EM6 &  \\
Observed events  (3.2 fb${}^{-1}$)& $1693$ & $874$     & $460$   & $349$      &  $111$  &  $40$    &  \\ 
SM prediction (post-fit)                     & $1693 \pm 41$ &   $874 \pm 30$ &   $460 \pm 21$ &   $349 \pm 19$ &   $111 \pm 11$ &   $40 \pm 6$  &  \\ 
SM prediction (pre-fit)           & $1880 \pm 150$ &   $971 \pm 79$ &   $488 \pm 40$ &  $439 \pm 36$ &   $131 \pm 12$ &   $50 \pm 5$  &  \\ \hline
\end{tabular*}
\end{footnotesize}
\end{center}
\label{tab:wen}
\end{table}

\begin{table}[!ht]
\caption{Data and SM background prediction, before and after the fit, in the $\wmn$ control region for the different selections.
For the SM predictions both the statistical and systematic uncertainties are included.
}
\begin{center}
\begin{footnotesize}
\begin{tabular*}{\textwidth}{@{\extracolsep{\fill}}lrrrrrrr}\hline
{\bf  Inclusive Selection}              & IM1 & IM2 & IM3 & IM4 & IM5 & IM6 & IM7 \\
Observed events  (3.2 fb${}^{-1}$)& $10481$ & $6279$ & $3538$ & $1939$ & $677$ & $261$ & $95$ \\ 
SM prediction (post-fit)                    & $10480 \pm 100$ &   $6279 \pm 79$ &   $3538 \pm 60$  &   $1939 \pm 44$ &   $677 \pm 26$ &   $261 \pm 16$ &   $95 \pm 10$ \\ 
SM prediction (pre-fit)   & $10500 \pm 710$ &   $6350 \pm 460$ &   $3560 \pm 280$ &   $2010 \pm 160$ &   $700 \pm 57$ &   $256 \pm 23$ &   $106 \pm 9$ \\
\hline
{\bf  Exclusive Selection}              & EM1 & EM2 & EM3 & EM4 & EM5 & EM6 & \\
Observed events  (3.2 fb${}^{-1}$)& $4202$ & $2741$ & $1599$ & $1262$ & $416$ & $166$ &  \\ 
SM prediction (post-fit)                     &  $4202 \pm 65$ &   $2741 \pm 52$ &   $1599 \pm 40$ &   $1262 \pm 36$ &   $416 \pm 20$ &   $166 \pm 13$ & \\ 
SM prediction (pre-fit) & $4140 \pm 260$ &   $2800 \pm 190$ &   $1540 \pm 120$ &   $1310 \pm 100$ &  $444 \pm 35$ &   $150 \pm 14$ & \\
\hline
\end{tabular*}
\end{footnotesize}
\end{center}
\label{tab:wmn}
\end{table}

\begin{table}[!ht]
\caption{Data and SM background prediction, before and after the fit, in the $\zmm$ control region for the different selections.
For the SM predictions both the statistical and systematic uncertainties are included.
}
\begin{center}
\begin{footnotesize}
\begin{tabular*}{\textwidth}{@{\extracolsep{\fill}}lrrrrrrr}\hline
{\bf  Inclusive Selection}              & IM1 & IM2 & IM3 & IM4 & IM5 & IM6 & IM7 \\
Observed events  (3.2 fb${}^{-1}$)& $1488$ & $877$ & $505$ & $293$ & $100$ & $33$ & $15$ \\ 
SM prediction (post-fit)                     & $1488 \pm 39$ &   $877 \pm 30$ &  $505 \pm 22$ &   $293 \pm 17$ &   $100 \pm 10$ &   $33 \pm 6$ &   $15 \pm 4$ \\ 
SM prediction (pre-fit)      & $1520 \pm 98$ &   $910 \pm 59$ &   $487 \pm 34$ &   $271 \pm 19$ &   $89 \pm 7$ &   $32 \pm 3$ &   $13 \pm 1$ \\
\hline
{\bf  Exclusive Selection}              & EM1 & EM2 & EM3 & EM4 & EM5 & EM6 &  \\
Observed events  (3.2 fb${}^{-1}$)& $611$ & $372$ & $212$ & $193$ & $67$ & $18$ &  \\ 
SM prediction  (post-fit)                    & $611 \pm 25$ &  $372 \pm 19$ &   $212 \pm 15$ &   $193 \pm 14$ &   $67 \pm 8$ &   $18 \pm 4$ &   \\ 
SM prediction  (pre-fit)  &  $610 \pm 42$ &   $422 \pm 36$ &   $217 \pm 15$ &   $182 \pm 13$ &   $57 \pm 4$ &   $19 \pm 2$ & \\ \hline
\end{tabular*}
\end{footnotesize}
\end{center}
\label{tab:zmm}
\end{table}

Figure~\ref{fig:cr1}
shows, for the IM1 monojet-like kinematic selection and in the different control regions, 
the distributions of the $\met$ and the 
leading-jet $\pt$ in data and MC simulation.  The MC predictions
include data-driven normalization factors as extracted from the global fit that considers 
exclusive $\met$ bins. Altogether, the MC  simulation provides a good 
description of the shape of the measured distributions in the different control regions.

%
%

\begin{figure}[!ht]
\begin{center}
\mbox{
  \includegraphics[width=0.45\textwidth]{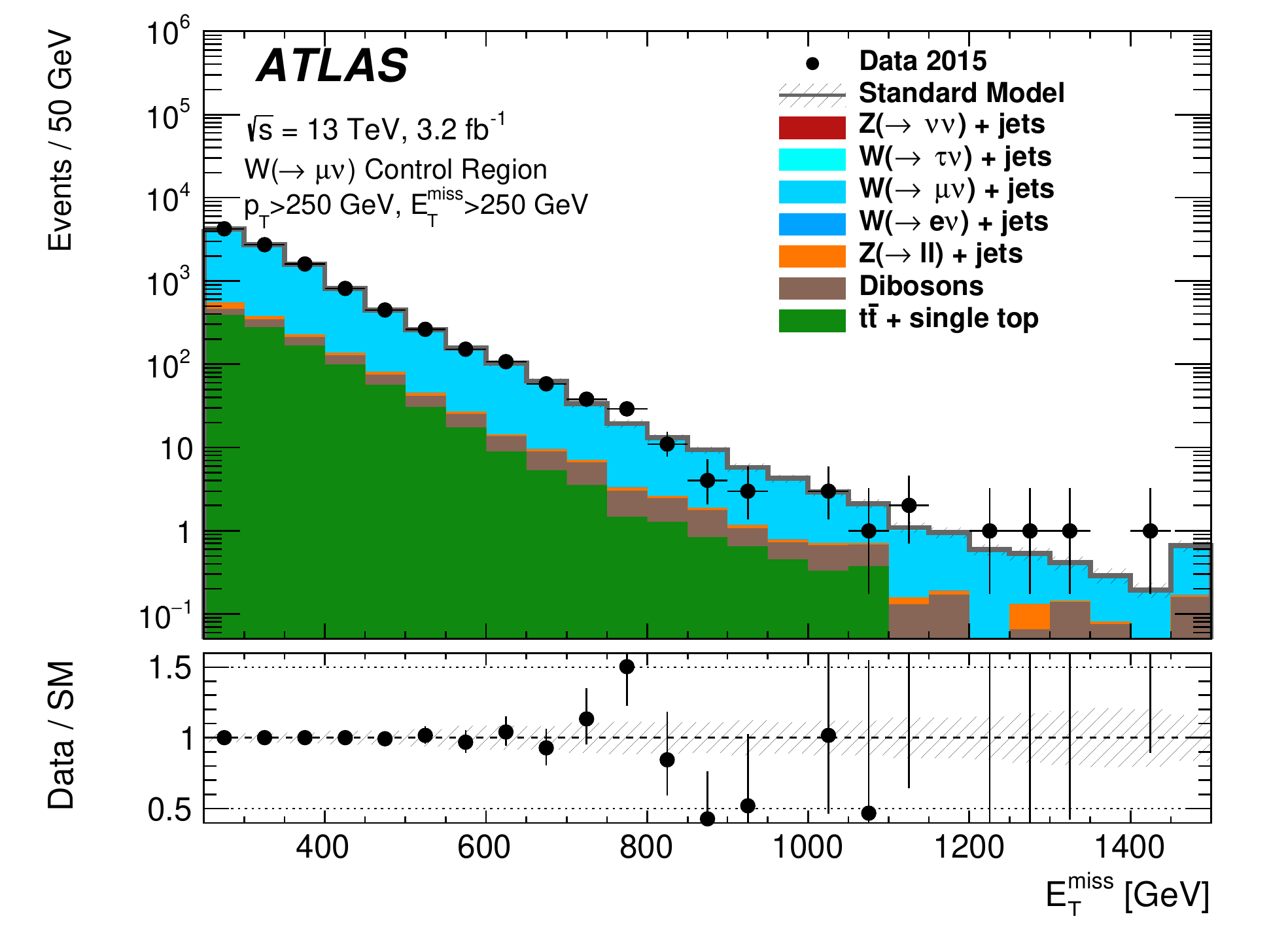}
  \includegraphics[width=0.45\textwidth]{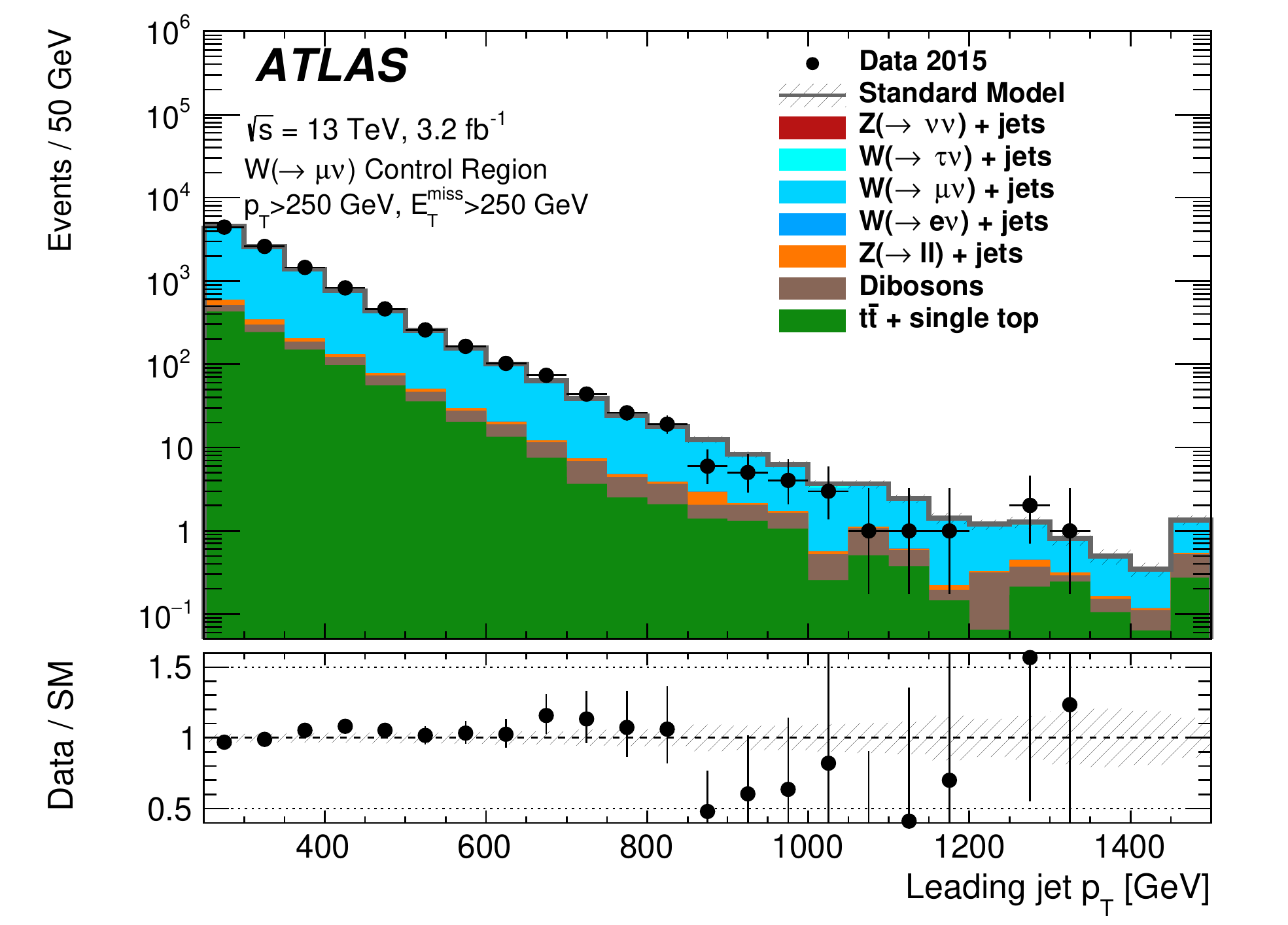}
}
\mbox{
  \includegraphics[width=0.45\textwidth]{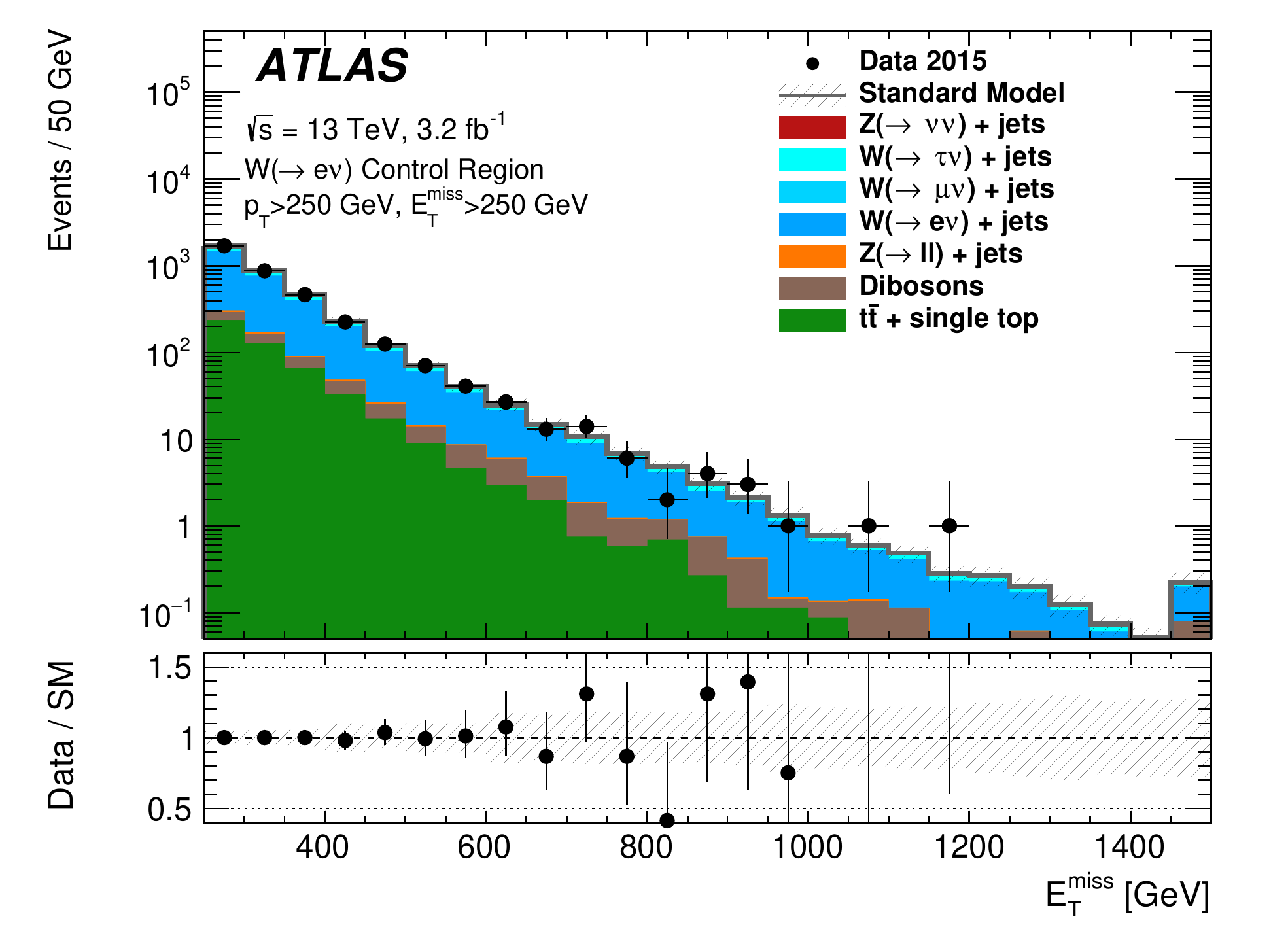}
  \includegraphics[width=0.45\textwidth]{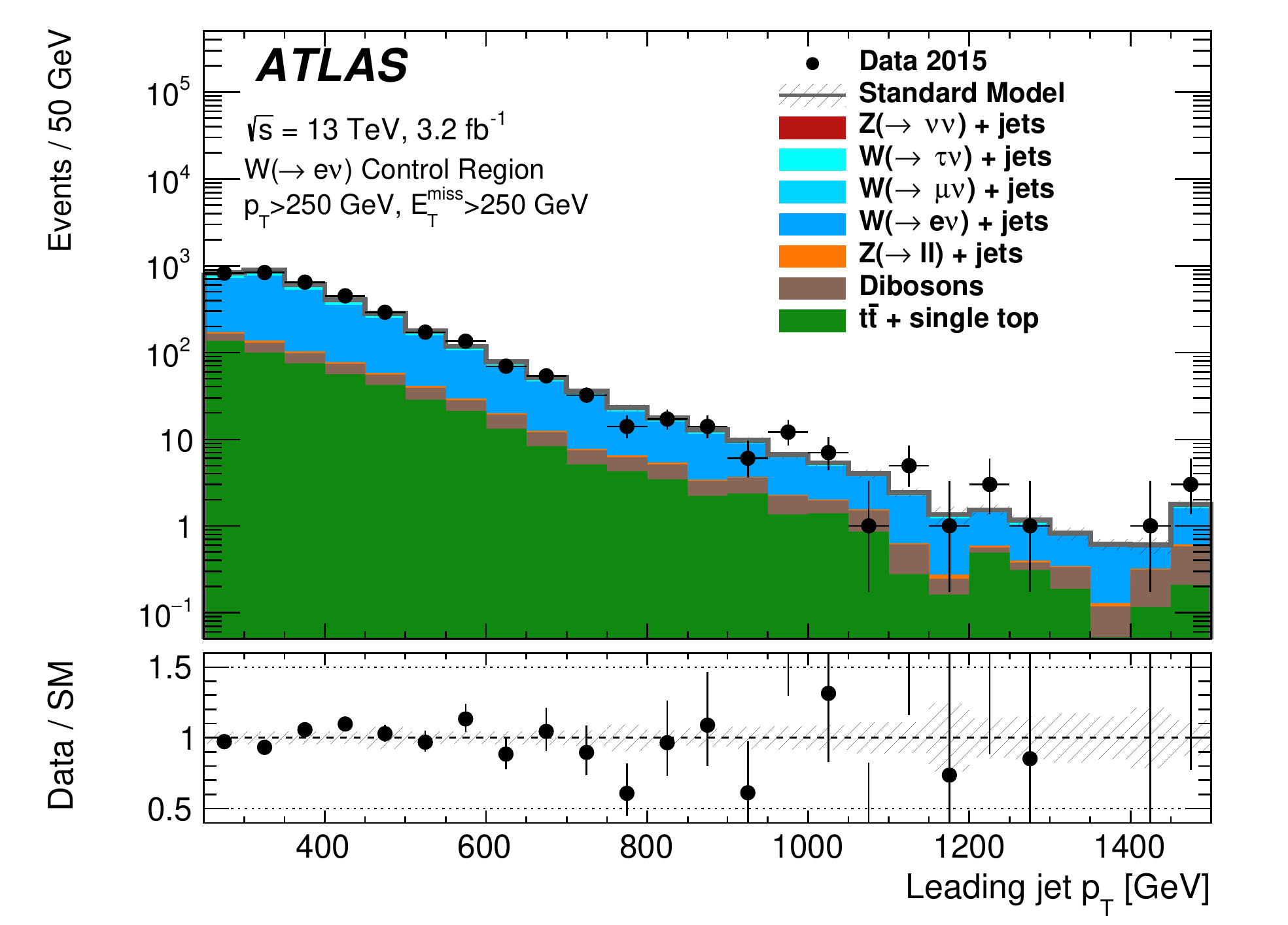}
}
\mbox{
  \includegraphics[width=0.45\textwidth]{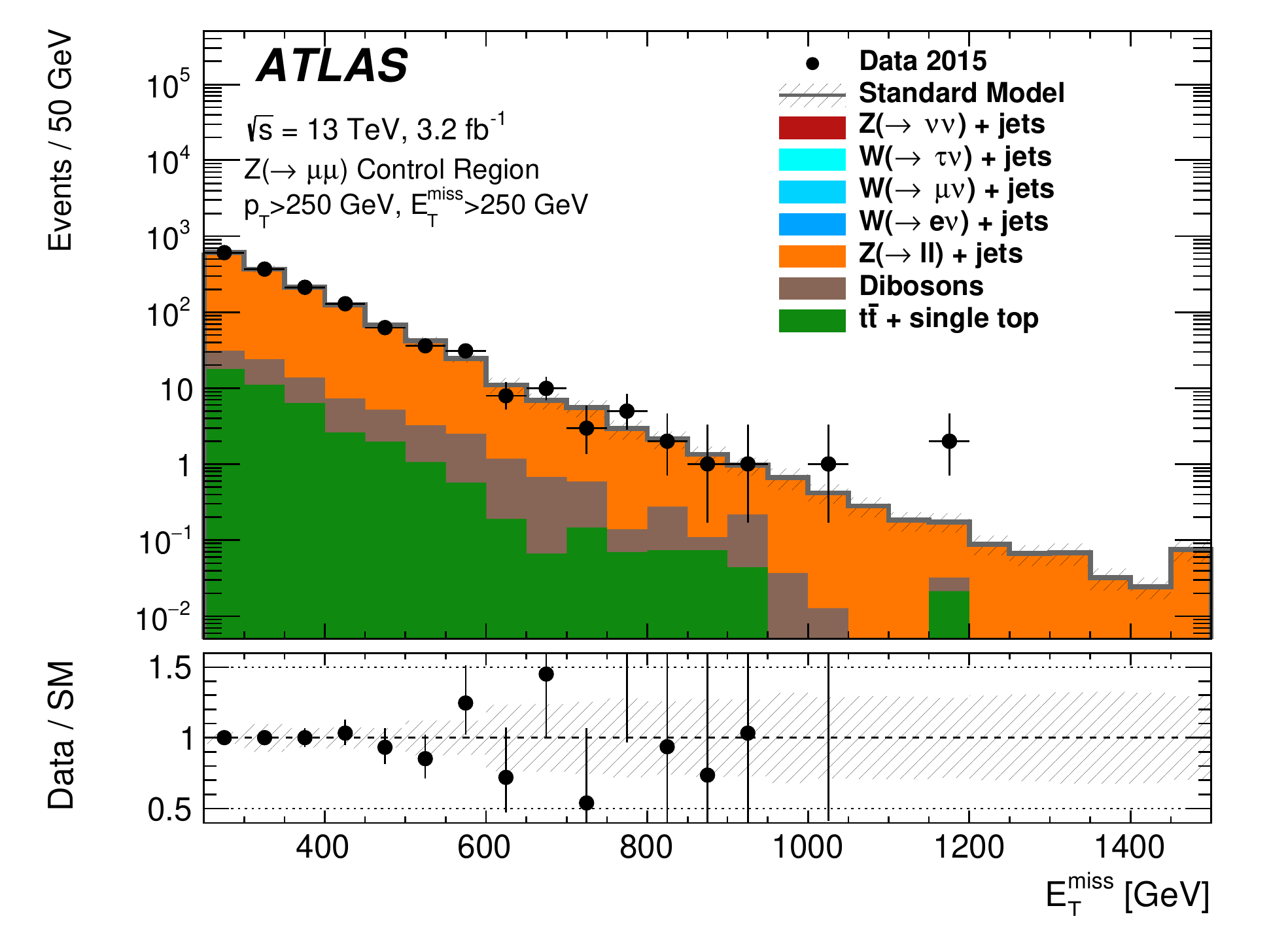}
  \includegraphics[width=0.45\textwidth]{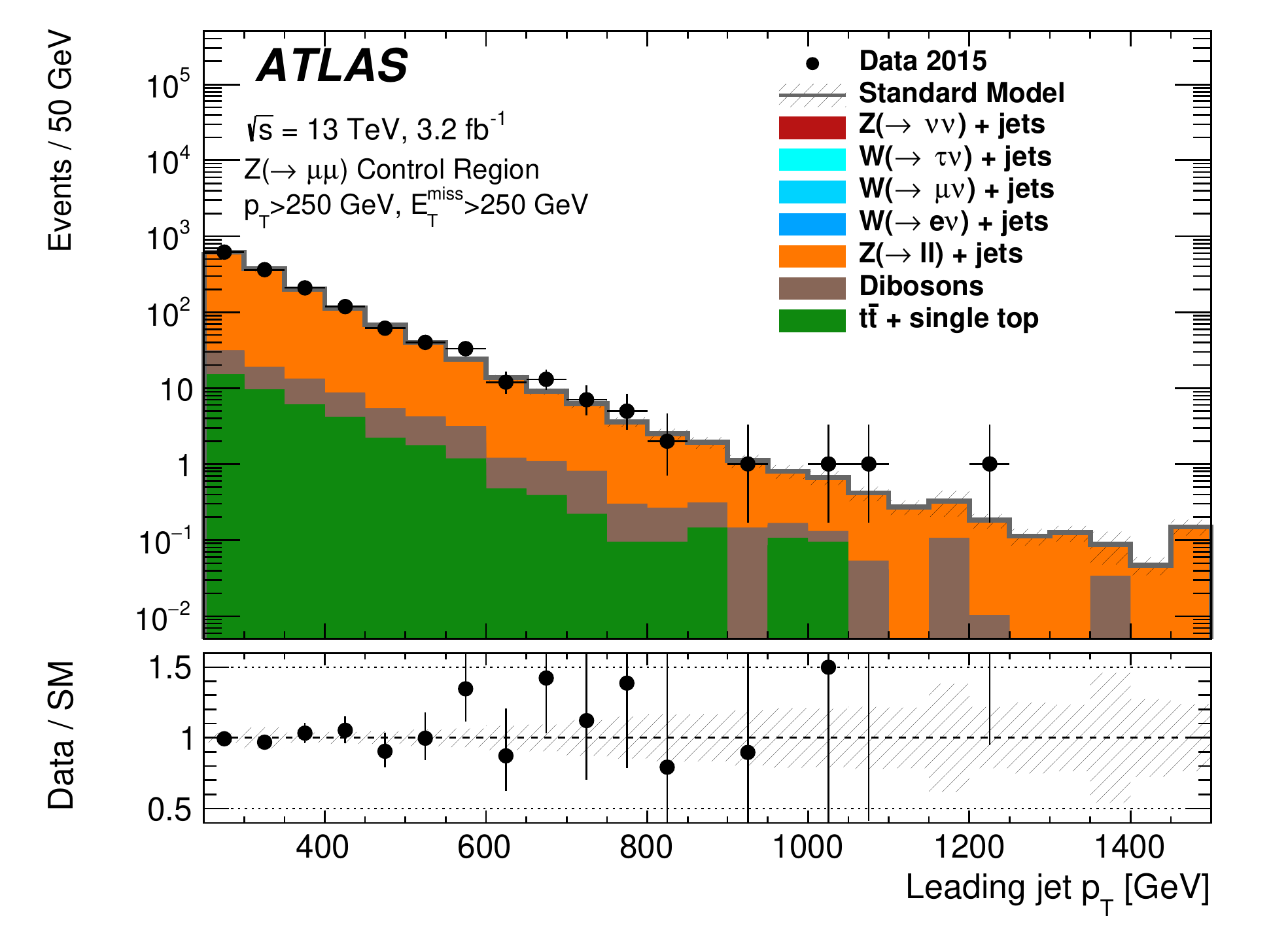}
}
\end{center}
\caption{
The measured $\met$ and leading-jet $\pt$ distributions
in the $\wmn$+jets (top),  $\wen$+jets (middle), and $\zmm$+jets (bottom) control regions, for the IM1 selection, compared to
the background predictions.
The latter include the global normalization factors extracted from the fit as performed in exclusive $\met$ bins.
The error bands in the ratios include the statistical and experimental  uncertainties in the background predictions as 
determined by the global fit to the data in the control regions.
The contributions from multijets and non-collision backgrounds are negligible and are not shown in the figures.
}
\label{fig:cr1}
\end{figure}

%
%

In the analysis, the control regions are defined using the same
requirements for $\met$, leading jet $\pt$, event topologies, and jet vetoes as in the
signal regions,  such that no extrapolation in $\met$ or jet $\pt$ is needed from control to signal regions.  
Agreement between data and background predictions is confirmed in a low-$\pt$ validation region defined using 
the same monojet-like  selection criteria with $\met$ limited to the range $150$--$250\GeV$.

 \section{Systematic uncertainties}
 \label{sec:syst}

In this section  the impact of each  source of systematic uncertainty on the total 
background prediction in the signal regions, as determined via the global fits 
explained in Section~\ref{sec:fitback}, is discussed. Here, the  case of the 
inclusive $\met$ selections is presented.  Similar studies are carried out in exclusive $\met$ bins.
The correlation of systematic uncertainties across $\met$ bins
is properly taken into account.
 Finally, the experimental and theoretical uncertainties in the 
 signal yields are discussed.

\subsection{Background systematic uncertainties}

%
%
Uncertainties in the absolute jet and $\met$ energy scales and resolutions~\cite{Aad:2011he}
translate  into an uncertainty in
the total background which varies between $\pm 0.5\%$ for IM1 and  $\pm 1.6\%$ for IM7.
Uncertainties related to jet quality requirements, pileup description and 
corrections to the jet $\pt$ and $\met$ introduce 
a $\pm 0.2\%$  to $\pm 0.9\%$ uncertainty in the background predictions.
%
%
Uncertainties in
the simulated lepton identification and reconstruction efficiencies, energy/momentum scale and resolution translate 
into an uncertainty in the total background which varies between
$\pm 0.1\%$ and $\pm 1.4\%$ for the IM1 and between $\pm 0.1\%$ and $\pm 2.6\%$ for the IM7 selections, respectively.

%
%

Variations of the renormalization, factorization, and parton-shower matching scales and PDFs 
in the  {SHERPA}  $W/Z$+jets background samples translate into a $\pm 1.1\%$ to $\pm 1.3\%$ uncertainty in the total background.
Model uncertainties, related to potential differences between $W$+jets and $Z$+jets final states, affecting the normalization of the 
dominant $\znn$+jets background and the small $\ztt$+jets background contribution as determined in $\wmn$+jets
and $\wen$+jets control regions,  are studied in detail. This includes uncertainties related to PDFs and  renormalization and factorization scale settings, the  
parton-shower parameters and the hadronization model used in the MC simulation, and the dependence on the lepton reconstruction and 
acceptance. As a result, an additional  $\pm 3\%$ uncertainty in the $\znn$+jets and $\ztt$+jets   contributions is included for all the selections.  
In addition, the effect from NLO electroweak corrections on the $W$+jets to $Z$+jets ratio is taken into account~\cite{Denner:2012ts,Denner:2011vu,Denner:2009gj}. 
Dedicated parton-level calculations are performed with the same $\met$ and leading-jet-$\pt$ requirements as  in the IM1--IM7
signal regions. The studies suggest an effect on the $W$+jets to $Z$+jets ratio which varies between about $\pm 1.9\%$ for IM1 and $\pm 5.2\%$ for IM7, although the 
calculations suffer from large uncertainties, mainly due to our limited knowledge of the photon PDFs in the proton. In this analysis, 
these results are  adopted as an additional uncertainty in the  $\znn$+jets and $\ztt$+jets contributions.
Altogether, this translates into an uncertainty in the total background which varies from  $\pm 2.0\%$ and $\pm 3.0\%$  for the IM1 and IM5 
selections, respectively, to about $\pm 3.9\%$ for the IM7 selection.

%
%

Theoretical uncertainties in the predicted background yields for top-quark-related processes include: uncertainties on 
the absolute $\ttbar$ and single-top production cross sections; variations in the set of parameters that govern the parton showers and the amount of initial- and final-state soft gluon radiation; and uncertainties 
due to the choice of renormalization and factorization scales and PDFs.  This introduces an uncertainty in the total background
prediction which varies between $\pm 2.7\%$ and $\pm 3.3\%$ for the IM1 and IM7 selections, respectively. 
Uncertainties in the diboson 
contribution are estimated using different MC generators and translate into an uncertainty in the total background in the 
range between $\pm 0.05\%$ and $\pm 0.4\%$.  
A $\pm 100\%$ uncertainty in the multijet and non-collision background 
 estimations  is adopted,  leading to a $\pm 0.2\%$ uncertainty in the total background for the IM1 selection.
Statistical uncertainties related to the data control regions and simulation samples lead to 
an additional uncertainty in the final background estimates in the signal regions  
which varies between $\pm 2.5\%$ for the IM1 and $\pm 10 \%$ for the IM7 selections. 
Finally, the impact of the uncertainty in the integrated luminosity, which partially cancels in the data-driven determination of the SM background, is negligible.


\subsection{Signal systematic uncertainties}
\label{sec:signalsys}

Several  sources of systematic uncertainty in the predicted signal yields are considered 
for each of the models of new physics. The uncertainties are computed separately for each signal region by  
varying the model parameters (see Section~\ref{sec:results}).

Experimental uncertainties include: those related to the jet and $\met$ reconstruction, energy
scales and resolutions; 
and the $\pm 5\%$ uncertainty in the integrated luminosity, derived following a methodology similar to that detailed in Ref.~\cite{Aad:2013ucp}, from a calibration of the luminosity scale using $x$--$y$ beam-separation scans performed in August 2015. 
Other uncertainties related to the jet quality requirements are negligible ($< 1\%$).

Uncertainties affecting the signal acceptance,
related to the generation of the signal samples, include:
uncertainties in the modeling of the initial- and final-state gluon radiation,
as determined using simulated samples 
with modified parton-shower parameters (by factors of two or one half) that enhance or suppress 
the parton radiation;
uncertainties due to PDF and variations of the $\alpha_{\rm s}(m_Z)$ value employed,  
as computed from the envelope of CT10, MMHT2014~\cite{Harland-Lang:2014zoa} and NNPDF30 error sets; 
and the choice of renormalization and factorization scales.
In addition, theoretical uncertainties in the predicted cross sections, including PDF and renormalization and factorization
scale  uncertainties, are computed separately for the different models.

\section{Results and interpretation}
\label{sec:results}

 The number of events in data and the expected background
predictions in several inclusive and exclusive signal regions, as 
determined using the global fit discussed in Section~\ref{sec:fitback},  are 
presented in detail in Table~\ref{tab:sr123}. The results for all the signal regions are summarized in Table~\ref{tab:srsum}.
Good agreement is observed between the data and the SM predictions in each case.
The  SM predictions for the inclusive selections are 
determined with a total uncertainty of $\pm 4.0\%$, $\pm 6.8\%$,  and $\pm 12\%$ for the IM1, IM5, and IM7 
signal regions, respectively, which include correlations between uncertainties in the individual 
background contributions. 

Figure~\ref{fig:sr1}  shows
several measured distributions  
compared to the SM predictions for $\met > 250\GeV$, for which 
the normalization factors applied to the MC predictions, and the related uncertainties,  
are determined  from the global fit carried out in exclusive $\met$ bins. 
For illustration  purposes, the distributions  include the impact of different ADD, SUSY, and WIMP scenarios.

\begin{figure}[!ht]
\begin{center}
\mbox{
\includegraphics[width=0.485\textwidth]{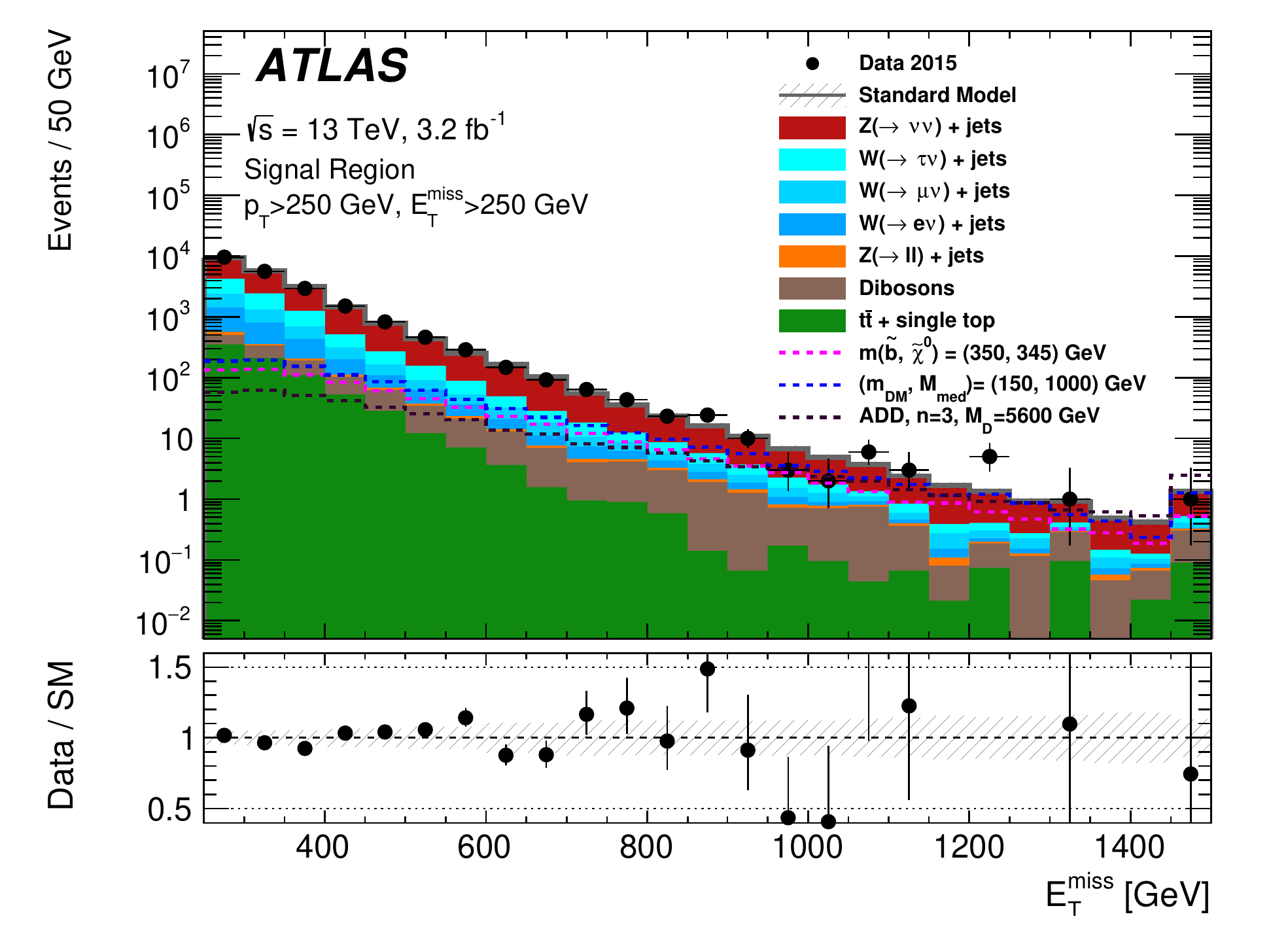}
\includegraphics[width=0.485\textwidth]{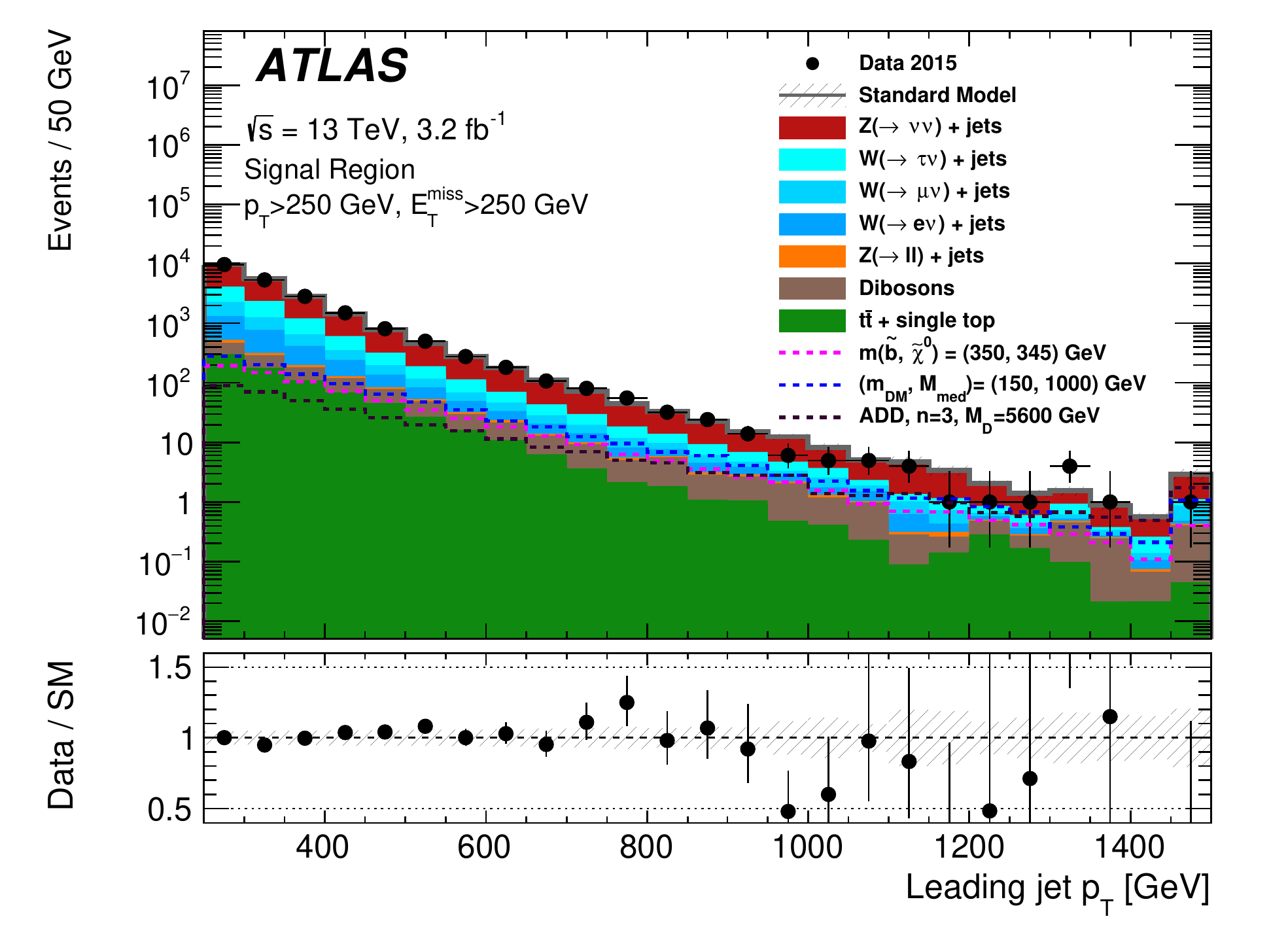}
}
\mbox{
\includegraphics[width=0.485\textwidth]{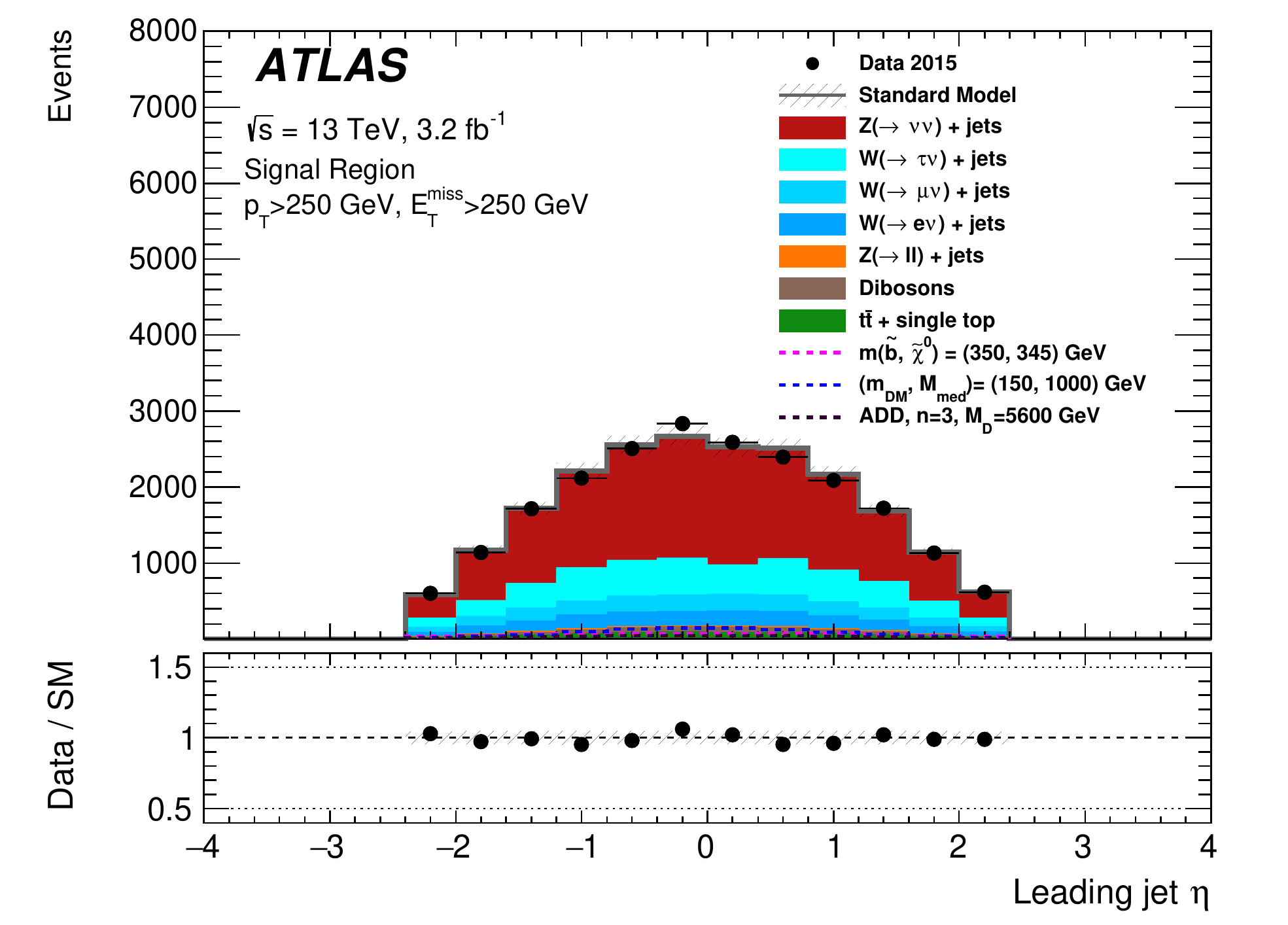}
\includegraphics[width=0.485\textwidth]{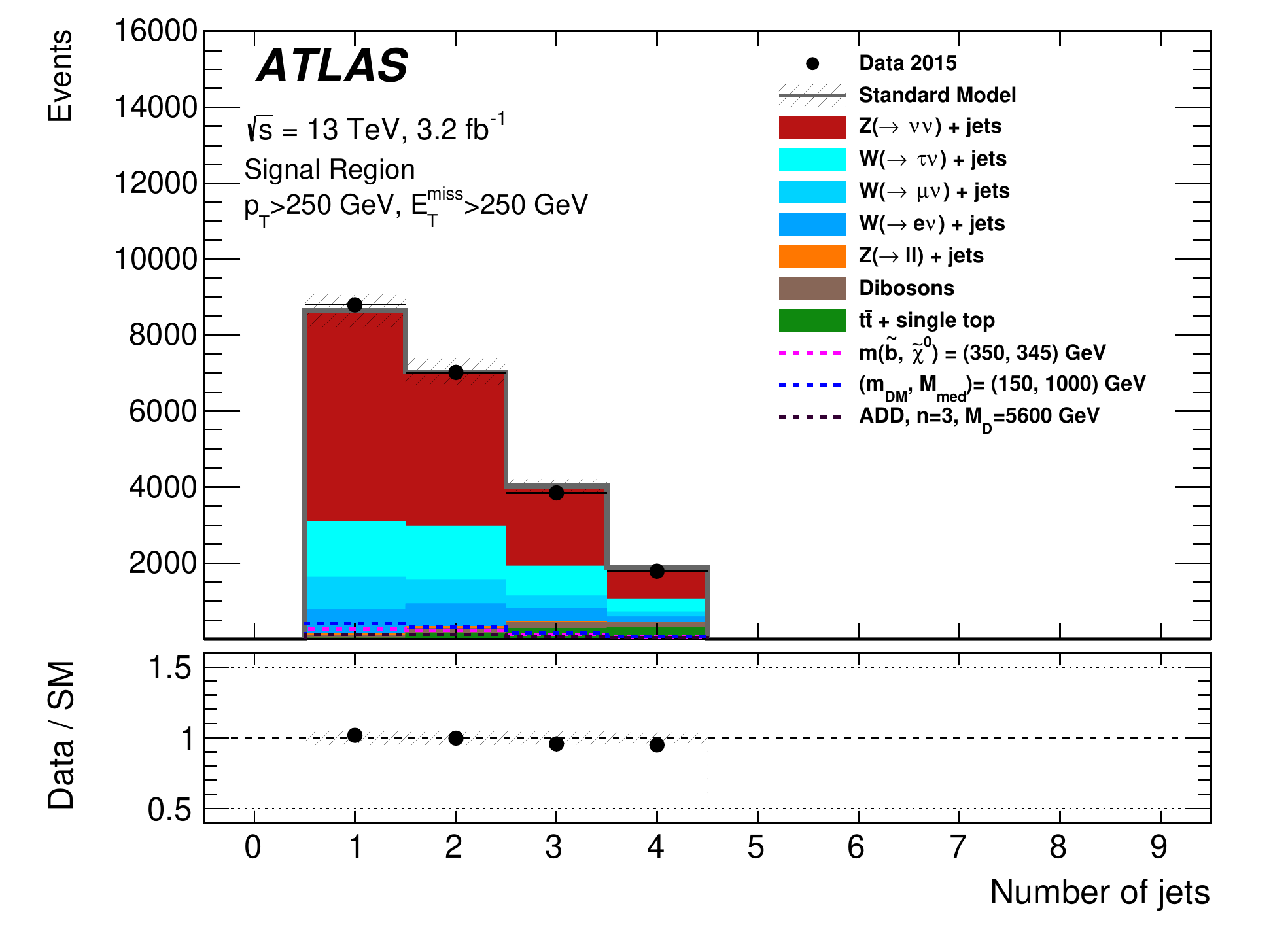}
}
\mbox{
\includegraphics[width=0.485\textwidth]{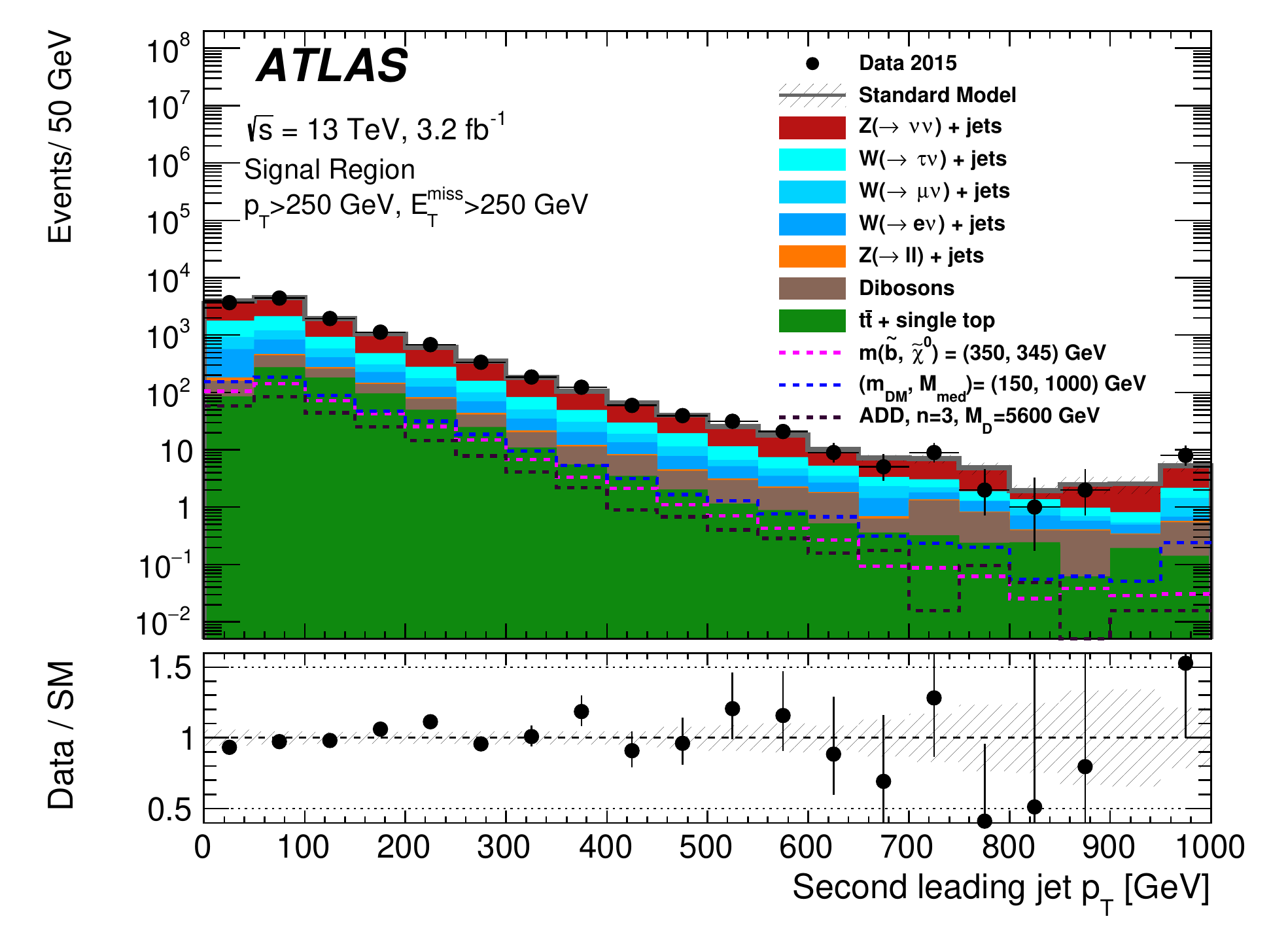}
\includegraphics[width=0.485\textwidth]{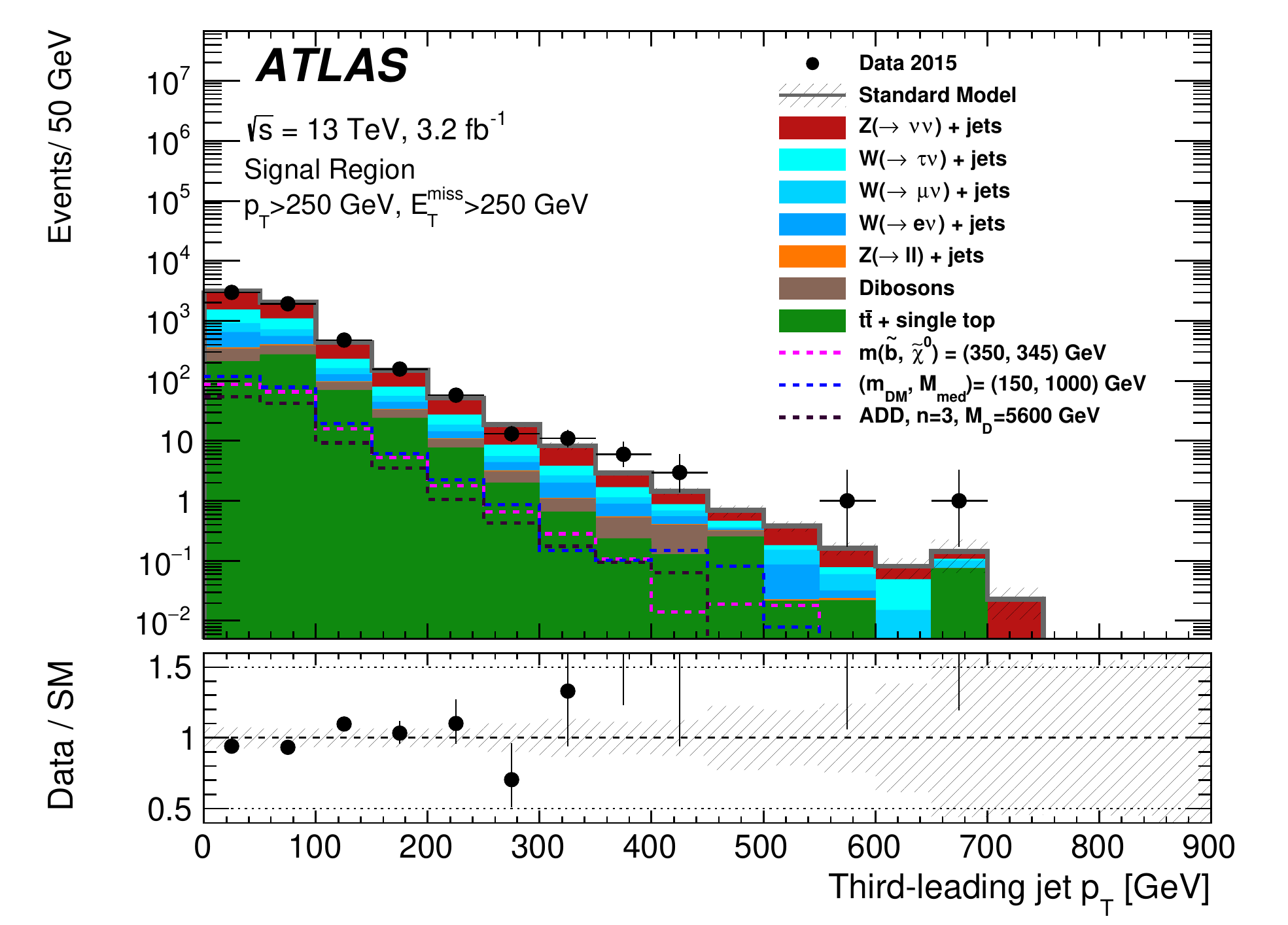}
}
\end{center}
\caption{
Measured distributions of the $\met$, leading-jet $\pt$, leading-jet $\eta$, jet multiplicity,  second-leading-jet $\pt$, and third-leading-jet $\pt$ for the  IM1 selection compared to the SM predictions.  The latter are normalized  with normalization 
factors as determined by the  global fit that considers exclusive $\met$ regions. 
For illustration purposes, the distributions of different ADD, SUSY, and WIMP scenarios  are included.
The error bands in the ratios shown in the lower panels include both the statistical and systematic uncertainties
in the background predictions.
The contributions from multijets and non-collision backgrounds are negligible and not shown in the figures.
}
\label{fig:sr1}
\end{figure}

\begin{table}[!ht]
\caption{Data and SM background predictions in the signal region for several inclusive and exclusive $\met$ selections.
For the SM prediction both the statistical and systematic uncertainties are included.
In each signal region, the individual uncertainties for the different background processes can be correlated,
and do not necessarily add in quadrature to the total background uncertainty.
}
\begin{center}
\begin{footnotesize}
\begin{tabular*}{\textwidth}{@{\extracolsep{\fill}}lrrrr}\hline
{\bf  Signal Region}               &  IM1            & EM3            & EM5           & IM7            \\
Observed events  (3.2 fb${}^{-1}$) & $21447$         & $2939$         & $747$         & $185$          \\ \hline
SM prediction                      & $21730 \pm 940$ & $3210 \pm 170$ & $686 \pm 50$  & $167 \pm 20$   \\ \hline
$\wen$                             & $1710 \pm 170$  & $228 \pm 26$   & $37 \pm 7$    & $7 \pm 2$      \\
$\wmn$                             & $1950 \pm 170$  & $263 \pm 28$   & $44 \pm 8$    & $11 \pm 2$     \\
$\wtn$                             & $3980 \pm 310$  & $551 \pm 47$   & $101 \pm 15$  & $19 \pm 4$     \\
$\zee$                             & $0.01 \pm 0.01$ & $-$            & $-$           & $-$            \\
$\zmm$                             & $76 \pm 30$   & $9 \pm 5$        & $5 \pm 2$     & $2 \pm 1$   \\
$\ztt$                             & $48 \pm 7$      & $5 \pm 1$      & $0.9 \pm 0.2$ & $0.2 \pm 0.1$  \\
$\znn$                             & $12520 \pm 700$ & $1940 \pm 130$ & $443 \pm 42$  & $109 \pm 18$   \\
$\ttbar$, single top               & $780 \pm 240$   & $108 \pm 32$   & $19 \pm 7$    & $3 \pm 1$      \\
Dibosons                           & $506 \pm 48$    & $82 \pm 8$     & $36 \pm 5$    & $15 \pm 2$     \\
Multijets                          & $51 \pm 50$     & $6 \pm 6$      & $1 \pm 1$ & $0.4 \pm 0.4$  \\ 
Non-collision background           & $110 \pm 110$   & $19 \pm 19$    & $-$           & $-$            \\ \hline\hline
\end{tabular*}
\end{footnotesize}
\end{center}
\label{tab:sr123}
\end{table}

\begin{table}[!ht]
\caption{Data and SM background predictions in the signal region for the different selections.
For the SM predictions both the statistical and systematic uncertainties are included.
}
\begin{center}
\begin{footnotesize}
\begin{tabular*}{\textwidth}{@{\extracolsep{\fill}}lrrrrrrr}\hline
{\bf  Signal Region}              & IM1   & IM2   & IM3   & IM4   & IM5   & IM6   & IM7   \\
Observed events  (3.2 fb${}^{-1}$)& $21447$   & $11975$   & $6433$   & $3494$   & $1170$   & $423$   & $185$   \\ 
SM prediction                     & $21730 \pm 940$ &  $12340 \pm 570$ &  $6570 \pm 340$ &  $3390 \pm 200$ &  $1125 \pm 77$ &  $441 \pm 39$ &  $167 \pm 20$  \\ \hline
{\bf  Signal Region}              & EM1   & EM2   & EM3   & EM4   & EM5   & EM6   &    \\
Observed events  (3.2 fb${}^{-1}$)& $9472$ & $5542$ & $2939$ & $2324$ & $747$ & $238$ &  \\ 
SM prediction                     & $9400 \pm 410$ &  $5770 \pm 260$ &  $3210 \pm 170$ &  $2260 \pm 140$ &  $686 \pm 50$ &  $271 \pm 28$ & \\ \hline
\end{tabular*}
\end{footnotesize}
\end{center}
\label{tab:srsum}
\end{table}

The level of  agreement between the data and the SM predictions for the total number of events  
in the different inclusive signal regions IM1--IM7 is translated into upper limits for the presence of 
new phenomena. 
A simultaneous likelihood fit is performed in both the control and signal regions,  separately for each 
of the inclusive regions IM1--IM7. As a result,
model-independent 95$\%$ confidence level (CL) upper limits 
on the visible cross section, defined as the production cross section times acceptance times 
efficiency $\sigma \times A \times \epsilon$, are extracted    
using the $CL_s$ modified frequentist approach~\cite{Read:2002hq} and 
considering  the systematic 
uncertainties in the SM backgrounds 
and the  uncertainty in the  integrated luminosity.
The results are presented in Table~\ref{tab:indep}. 
Values of $\sigma \times A \times \epsilon$  
above 553~fb (for IM1) and above 19~fb (for IM7)  are excluded at 95$\%$ CL.  
Typical event selection efficiencies $\epsilon$  varying from about
$100\%$ for IM1 to  $96\%$ for IM7 are found in simulated 
$\znn$+jets background processes.

\begin{table}[!ht]
\caption{
Observed and expected 95$\%$ CL upper limits on the  
number of signal events, $S_{\rm obs}^{95}$ and  $S_{\rm exp}^{95}$, and on the                                                              
visible cross section, defined as the product of cross section, acceptance and efficiency,  
$\langle{\rm \sigma}\rangle_{\rm obs}^{95}$,  
for the IM1--IM7 selections.}             
\begin{center}
\setlength{\tabcolsep}{0.0pc}
{\footnotesize
\begin{tabular*}{\textwidth}{@{\extracolsep{\fill}}lrrr}
\noalign{\smallskip}\hline\noalign{\smallskip}
{\bf Signal channel}                        & $\langle{\rm \sigma}\rangle_{\rm obs}^{95}$~[fb]  &  $S_{\rm obs}^{95}$  & $S_{\rm exp}^{95}$ \\ 
\midrule
IM1   & $553$ &  $1773$  & ${1864}^{+829}_{-548}$ \\
IM2   & $308$ &  $988$  & ${1178}^{+541}_{-348}$ \\
IM3   & $196$ &  $630$  & ${694}^{+308}_{-204}$ \\
IM4   & $153$ &  $491$  & ${401}^{+168}_{-113}$ \\
IM5   & $61$ &  $196$  & ${164}^{+63}_{-45}$ \\
IM6   & $23$ &  $75$  & ${84}^{+32}_{-23}$ \\
IM7   & $19$ &  $61$  & ${48}^{+18}_{-13}$ \\
\noalign{\smallskip}\hline\noalign{\smallskip}
\end{tabular*}
}
\end{center}
\label{tab:indep}
  \end{table}

\subsection{Large extra spatial dimensions}

The level of agreement between the data and the SM predictions is translated into limits  on the parameters of the ADD model.
Only the signal regions with $\met > 400\GeV$, 
where the SM background is moderate and the shape difference between signal and the SM background becomes apparent, 
have an impact on the ADD limits.
The typical $A \times \epsilon$ of the selection criteria varies, as the number of extra dimensions $n$ increases from $n=2$ to $n=6$, 
between 5.5$\%$ and 6.6$\%$ for IM4  and between $2.9\%$ and 4.2$\%$ for IM7.

The experimental uncertainties related to the jet and $\met$
scales and resolutions introduce uncertainties in the signal yields which vary between
$\pm 1\%$ and $\pm 3\%$.
The uncertainties related to the modeling of the initial- and final-state 
gluon radiation  
translate into 
uncertainties in the ADD signal acceptance which  
vary between $\pm 7\%$ and $\pm 10\%$.
The  uncertainties due to the
PDFs, affecting the predicted signal cross sections, increase from $\pm 16\%$ at $n=2$ to $\pm 42\%$ at $n=6$.
The effect of PDF uncertainties on the acceptance is between $\pm 10\%$ and $\pm 20\%$, 
mildly increasing with increasing $n$ and $\met$.
Similarly, the variations of the renormalization and factorization scales 
introduce a $\pm 23\%$ to $\pm 36\%$ uncertainty 
in the signal yields, with increasing $n$ and $\met$ requirements, 
and about a 
$\pm 10\%$ variation in the signal acceptance.   

Observed and expected 95$\%$ CL exclusion limits are set on $M_D$ as a function of $n$
using the $CL_s$ approach, for which a simultaneous fit to the signal and control regions  
in the exclusive $\met$ bins  is performed,  
including statistical and systematic uncertainties. 
Uncertainties in the signal acceptance times efficiency, 
the background predictions, and the luminosity are considered, and 
correlations between systematic uncertainties in signal and background predictions 
are taken into account. The fit accounts for 
the contamination of the control regions by signal events
which a priori is estimated to be very small.
In addition, observed limits are computed 
using 
$\pm 1\sigma$ variations of the theoretical predictions for the ADD  cross sections.
The $- 1\sigma$ variations of the ADD
theoretical cross sections result in about a 6$\%$ decrease in the nominal observed limits.
Figure~\ref{fig:add} and Table~\ref{tab:add} present the results in the case of the ADD model.
Values of $M_D$ below $6.58\TeV$ at $n=2$ and below $4.31\TeV$ at $n=6$ are excluded at
95$\%$~CL,  
which extend the exclusion from previous results using $8\TeV$ data~\cite{Aad:2015zva}. 

As discussed in Refs.~\cite{ATLAS:2012ky, Aad:2015zva}, the  analysis partially probes the phase-space region with $\hat{s} > M_D^2$, where $\sqrt{\hat{s}}$ is the
center-of-mass energy of the hard interaction. This challenges the validity of model implementation and the lower bounds
on $M_D$, as they depend on  the unknown ultraviolet behavior of the effective theory.  
The observed 95$\%$ CL limits are recomputed after 
suppressing, with a weighting  factor $M_D^4/\hat{s}^2$, the signal events with $\hat{s} > M_D^2$, here 
referred to as damping.  This results in a decrease of the quoted 95$\%$  CL lower limit on $M_D$  which is negligible for $n=2$ and about 
5$\%$ for $n=6$.

\begin{table}[ht!]
   \caption{The $95\%$ CL observed and expected lower 
limits on the fundamental Planck scale in $4+n$ dimensions, $M_D$, as a function of the number of extra dimensions $n$,  considering nominal LO signal cross sections.
The impact
of the $\pm 1\sigma$   theoretical uncertainty on
the observed limits and the  expected $\pm 1\sigma$ range of limits in the absence of a
signal are also given. Finally, the $95\%$ CL observed limits after damping of the signal cross section for  $\hat{s} > M_D^2$ 
(see text)
are quoted in parentheses.
}
\begin{center}
\begin{footnotesize}
\begin{tabular*}{\textwidth}{@{\extracolsep{\fill}}lcccc}\hline
  \multicolumn{5}{c}{$95\%$ CL  lower limits on $M_D$ [\TeV]} \\ \hline  
   $n$ extra   & \multicolumn{2}{c}{$95\%$ CL observed limit}     &  \multicolumn{2}{c}{$95\%$ CL expected limit}           \\ 
dimensions & Nominal (Nominal after damping) & $\pm$ 1$\sigma$ (theory) & Nominal & $\pm$ 1$\sigma$ (expected) \\ \hline
\noalign{\smallskip}
2 & 6.58 (6.58) & $^{+0.52}_{-0.42}$ & 6.88 & $^{+0.65}_{-0.64}$\\
\noalign{\smallskip}
3 & 5.46 (5.44) & $^{+0.45}_{-0.34}$ & 5.67 & $^{+0.41}_{-0.41}$\\
\noalign{\smallskip}
4 & 4.81 (4.74) & $^{+0.41}_{-0.29}$ & 4.96 & $^{+0.29}_{-0.29}$\\
\noalign{\smallskip}
5 & 4.48 (4.34) & $^{+0.41}_{-0.26}$ & 4.60 & $^{+0.23}_{-0.23}$\\
\noalign{\smallskip}
6 & 4.31 (4.10) & $^{+0.41}_{-0.24}$ & 4.38 & $^{+0.19}_{-0.19}$\\
\noalign{\smallskip}
   \hline
   \hline
\end{tabular*}
  \end{footnotesize}
\end{center}
\label{tab:add}
\end{table}

\begin{figure}[h]
\begin{center}
  \includegraphics[width=0.5\textwidth]{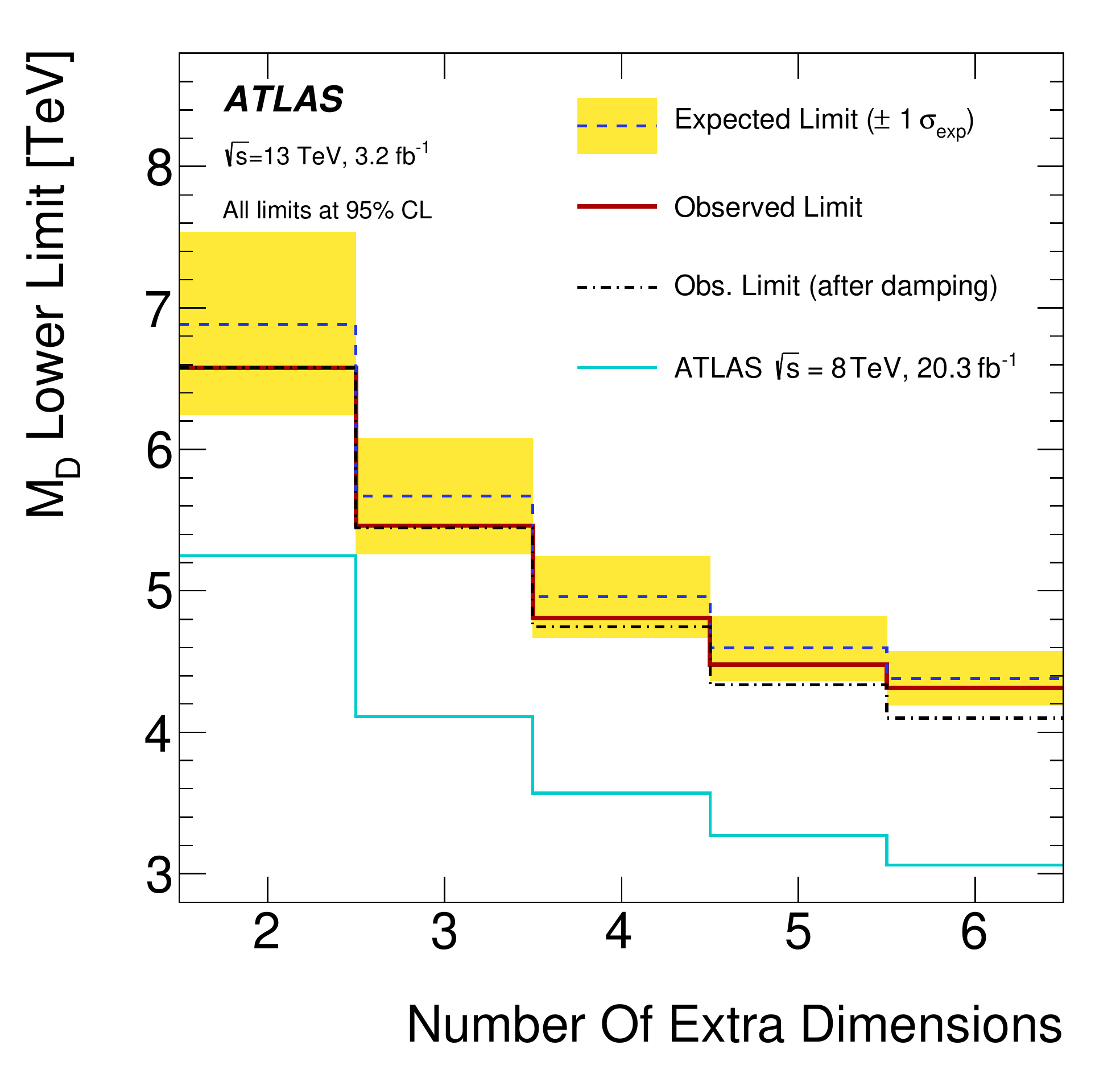}
\end{center}
\caption{
Observed and expected 95$\%$ CL lower limits on the fundamental Planck scale in $4+n$ dimensions, $M_D$, as a function of the number 
of extra dimensions. 
The  shaded area around the 
expected limit  indicates the expected $\pm 1\sigma$ 
range of limits in the absence of a signal. Finally, the thin dashed line shows the 95$\%$ CL observed limits 
after the suppression  of the events with 
$\hat{s} > M_D^2$ (damping) is applied,  as described in the text.  
The results from this analysis are compared to previous results from the ATLAS Collaboration at $\sqrt{s} = 8\TeV$~\cite{Aad:2015zva}.
}
\label{fig:add}
\end{figure}

%
%

\subsection{Squark pair production}

The results are translated into exclusion limits computed separately  for stop pair production    
with $\tilde{t}_1 \to  c +  \tilde{\chi}^{0}_{1}$, squark pair production with $\tilde{q} \to q + \ninoone$ ($q =u,d,c,s$), and sbottom pair production with $\tilde{b}_1 \to b + \ninoone$,  
as a function of the squark mass for different neutralino masses.   
As an example, in the case of stop pair production 
the typical $A \times \epsilon$ of the selection criteria varies, with increasing stop and neutralino masses, 
between 0.7$\%$ and 1.4$\%$ for IM1  and between $0.06\%$ and 0.8$\%$ for IM7.
Observed and expected 95$\%$ CL exclusion limits are calculated  
using a simultaneous fit to the signal and control regions in exclusive $\met$ bins, as in the case of the ADD models. 

The systematic uncertainties in the SUSY signal yields are also determined following a procedure close to that for 
the ADD case. The uncertainties related to the jet and $\met$
scales and resolutions introduce uncertainties in the signal yields which vary between $\pm 0.2\%$ and $\pm 7\%$
for different selections and  squark and neutralino masses.  
In addition,  the 
uncertainty in the integrated luminosity is included.
The uncertainties related to the modeling of 
initial- and final-state gluon radiation   
translate into a $\pm 7\%$ to $\pm 17\%$ 
uncertainty in the signal yields. 
The uncertainties due to the PDFs 
result in a $\pm 5\%$ to $\pm 17\%$ uncertainty in the signal yields.
Finally, the variations of the renormalization and factorization scales 
introduce a $\pm 4\%$ to $\pm 13\%$ uncertainty in the signal yields.

%
%

Figure~\ref{fig:stop} presents the results in the case of the $\tilde{t}_1 \to  c +  \tilde{\chi}^{0}_{1}$ signal. 
The previous limits from the ATLAS Collaboration~\cite{Aad:2014nra} are also shown. 
As anticipated, the monojet-like selection improves significantly the sensitivity at very low $\Delta m$.
In the compressed scenario with the stop and neutralino nearly degenerate in mass, the exclusion extends up to stop masses of 
$323\GeV$.  The region with $\Delta m < 5\GeV$ is not considered in the exclusion since 
in this regime the stop could become long-lived.  
Figure~\ref{fig:sbottom}~(left) presents the  observed  and expected 95$\%$ CL exclusion limits
as a function of the sbottom mass  and the sbottom--neutralino mass difference for the  $\tilde{b}_1 \to  b + \tilde{\chi}^{0}_{1}$ decay channel. 
In the scenario with $m_{\tilde{b}_1} - m_{\ninoone} \sim m_b$, this analysis extends the 95$\%$ CL
exclusion limits up to a sbottom mass of $323\GeV$.
Similarly, Fig.~\ref{fig:sbottom}~(right) presents the observed and expected 95$\%$ CL exclusion limits 
as a function of the squark mass and the squark--neutralino mass difference for $\tilde{q} \to q + \ninoone$ ($q =u,d,c,s$). 
In the compressed scenario with similar squark and neutralino masses, squark masses below $608\GeV$ are excluded at 95$\%$ CL.
These results significantly extend previous 
exclusion limits~\cite{Aaltonen:2012tq,Abazov:2008rc,Aad:2014nra}. 

\begin{figure}[h]
\begin{center}
\mbox{
  \includegraphics[width=0.8\textwidth]{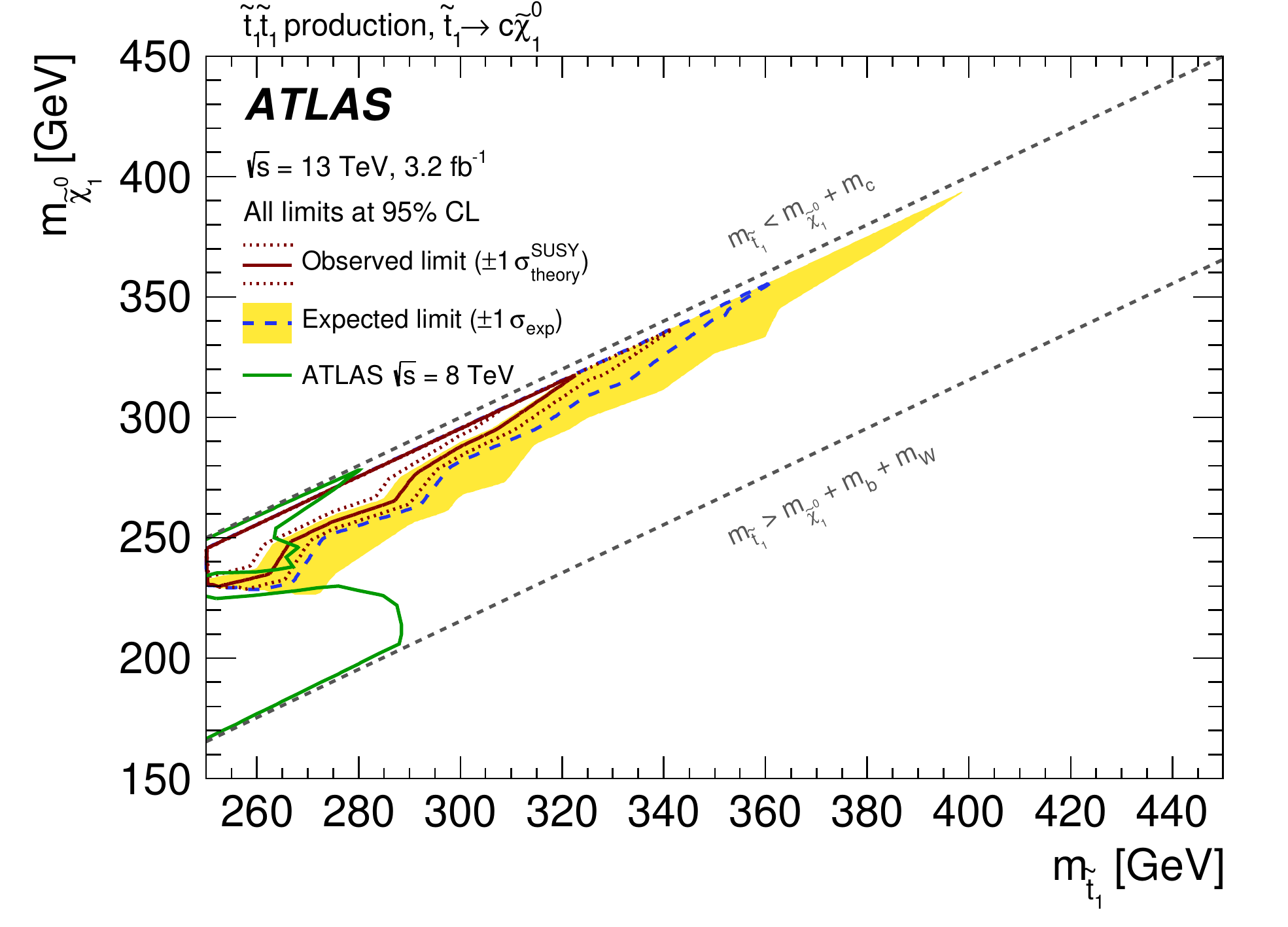}
}
\end{center}
\caption{
Excluded region at the 95$\%$ CL in the ($\tilde{t}_1,\tilde{\chi}^{0}_{1}$) mass plane for the decay channel  $\tilde{t}_1 \to  c +  \tilde{\chi}^{0}_{1}$ (BR = 100$\%$). 
The dotted lines around the observed limit
indicate the range of observed limits corresponding to $\pm 1\sigma$ variations of the NLO SUSY cross-section predictions.  The 
shaded area around the 
expected limit  indicates the expected $\pm 1\sigma$
ranges of limits in the absence of a signal. 
The results from this analysis are compared to previous results from the ATLAS Collaboration at $\sqrt{s} = 8\TeV$~\cite{Aad:2014nra}.
}
\label{fig:stop}
\end{figure}

\begin{figure}[h]
\begin{center}
\mbox{
  \includegraphics[width=0.48\textwidth]{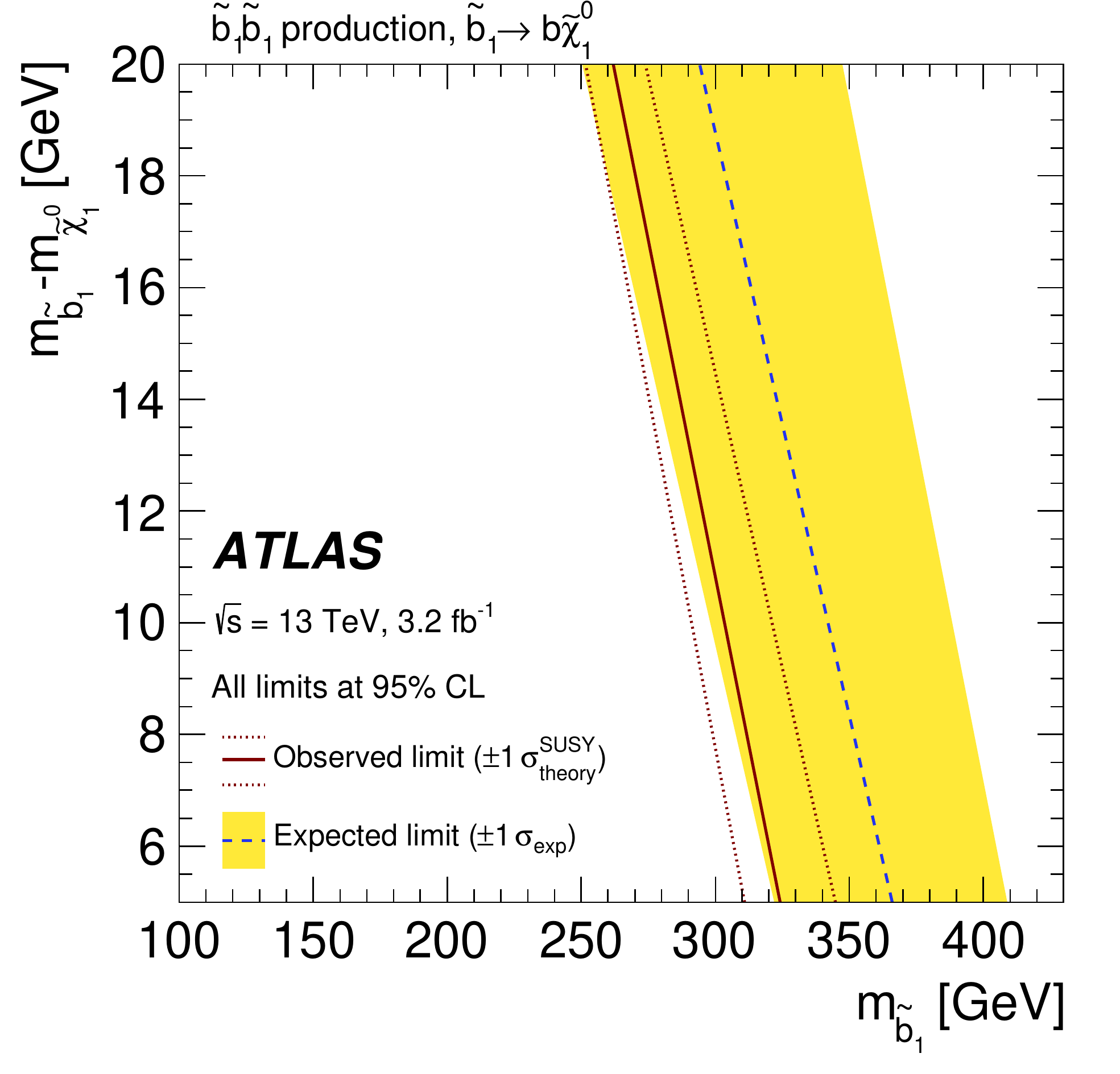}
  \includegraphics[width=0.48\textwidth]{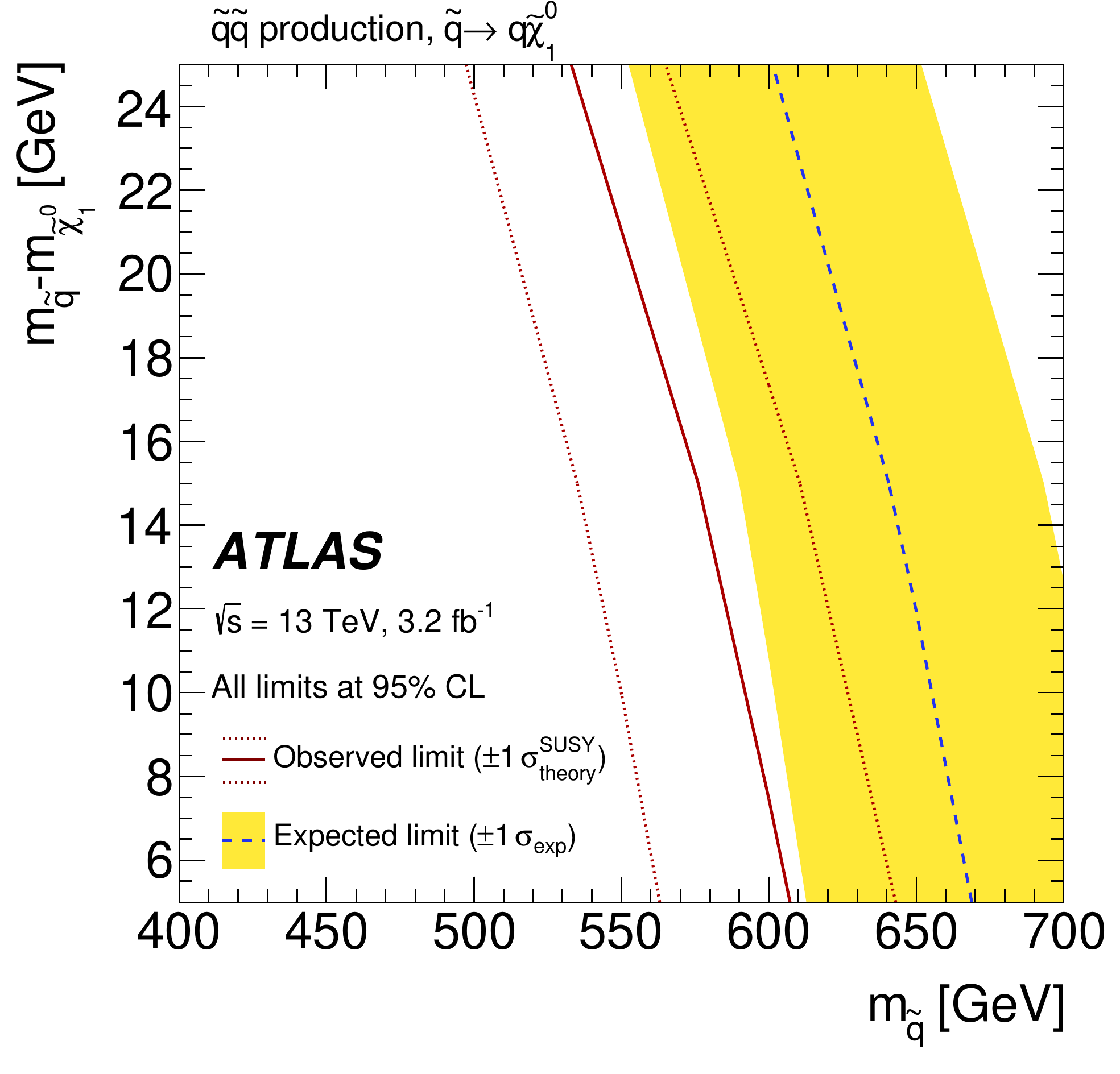}
}
\end{center}
\caption{
Exclusion region at 95$\%$ CL as a function of squark mass  and the squark--neutralino mass difference for (left) the decay channel  
$\tilde{b}_1 \to  b +  \tilde{\chi}^{0}_{1}$  and (right) $\tilde{q} \to  q +  \tilde{\chi}^{0}_{1}$ ($q =u,d,c,s$).
The dotted lines around the observed limit
indicate the range of observed limits corresponding to $\pm 1\sigma$ variations of the NLO SUSY cross-section predictions.  The 
shaded area around the 
expected limit  indicates the expected $\pm 1\sigma$
ranges of limits in the absence of a signal.
} 
\label{fig:sbottom}
\end{figure}

\subsection{Weakly interacting massive particles}

The results are translated into exclusion limits on the WIMP pair-production, 
assuming the exchange of an axial-vector mediator in the $s$-channel.
For  on-shell WIMP pair-production, where $m_A>2m_\chi$, typical $A \times \epsilon$  values for the signal models  with a 
$1\TeV$ mediator range 
from 25\%  to 2\% for IM1 and  IM7 selections, respectively.

The effect of experimental uncertainties related to jet and $\met$ 
scales and resolutions is found to be similar to 
the effect in the ADD model. 
The uncertainty related to the modeling of the initial- and final-state radiation translates into $\pm 20\%$ uncertainty in the acceptance and is neglected for the cross section.
The choice of different PDF sets results in up to $\pm 20\%$ uncertainty in the acceptance and $\pm 10\%$ uncertainty in the cross section. 
Varying the
renormalization and factorization scales 
introduces $\pm 5\%$ variations of the cross section and a $\pm 3\%$ change in the acceptance.
In addition,  the
uncertainty in the integrated luminosity is included.

Figure~\ref{fig:dm}~(left) shows the observed and expected 95\% CL exclusion limits in the $m_\chi$--$m_A$ parameter plane
for a simplified model with an axial-vector mediator,
Dirac WIMPs, and couplings
$g_q = 1/4$ and $g_\chi = 1$. A minimal mediator width is assumed. 
In addition, observed limits are shown using $\pm1\sigma$ theoretical  uncertainties  in the signal cross sections.
In the on-shell regime, the models with mediator masses up to $1\TeV$ are excluded. This analysis loses sensitivity to the models in the off-shell regime, where the decay into a pair of WIMPs is kinematically suppressed. 
The perturbative unitarity is violated in the parameter region defined by $m_\chi > \sqrt{\pi/2}\ m_A$ \cite{Kahlhoefer:2015bea}. 
The masses corresponding to the correct relic density as measured by the Planck and WMAP satellites~\cite{Adam:2015rua,2013ApJS..208...19H}, in the absence of any interaction other than the one considered, are indicated in the figure as a line that crosses the excluded region at $m_A \sim 880$~GeV and $m_\chi \sim 270$~GeV.
The region towards lower WIMP masses or higher mediator masses corresponds to dark matter overproduction. On the opposite side of the curve, other WIMP production mechanisms need to exist in order to explain the observed dark matter relic density.

In Fig.~\ref{fig:dm}~(right) the results are translated into 90\% CL exclusion limits on the spin-dependent WIMP--proton scattering cross section as a function of the WIMP mass, following the prescriptions explained in Refs.~\cite{Buchmueller:2014yoa,Boveia:2016mrp}, and are compared to results from the direct-detection experiments XENON100 \cite{Aprile:2013doa}, LUX \cite{Akerib:2016lao}, and PICO \cite{Amole:2015pla,Amole:2016pye}.
This comparison is model-dependent and solely valid in the context of this particular $Z'$-like model. 
In this case, stringent limits on the scattering cross section of the order of $10^{-42}$\,cm$^2$ up to WIMP masses of about $300\GeV$ are inferred from this analysis, and complement the results from direct-detection experiments for $m_\chi < 10\GeV$.
The loss of sensitivity in models where WIMPs are produced off-shell is expressed by the turn of the exclusion line, reaching 
back to low WIMP masses and intercepting the exclusion lines from the direct-detection experiments at around $m_\chi=80\GeV$.  

\begin{figure}
\centering
\includegraphics[width=0.45\textwidth]{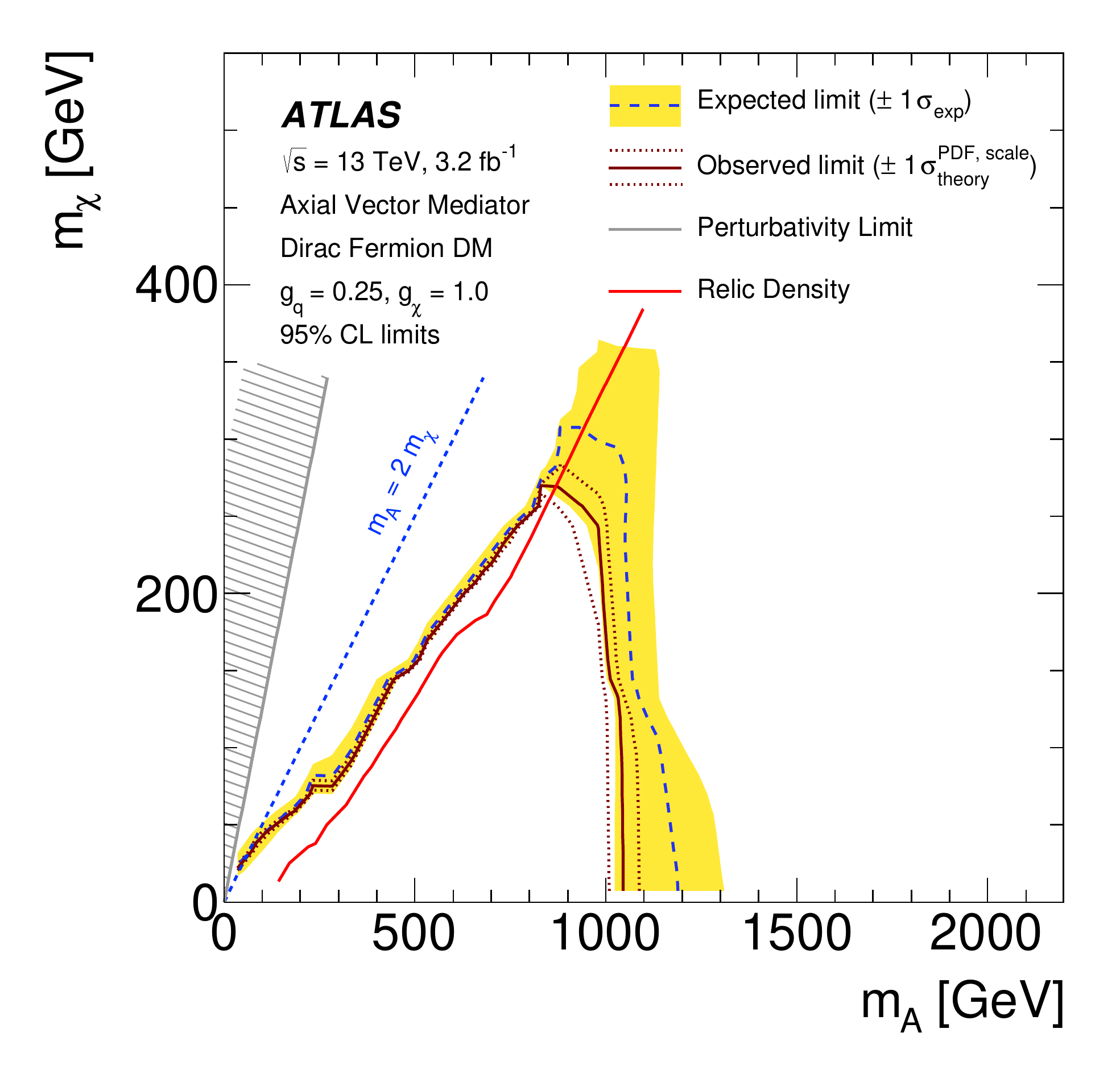}
\includegraphics[width=0.45\textwidth]{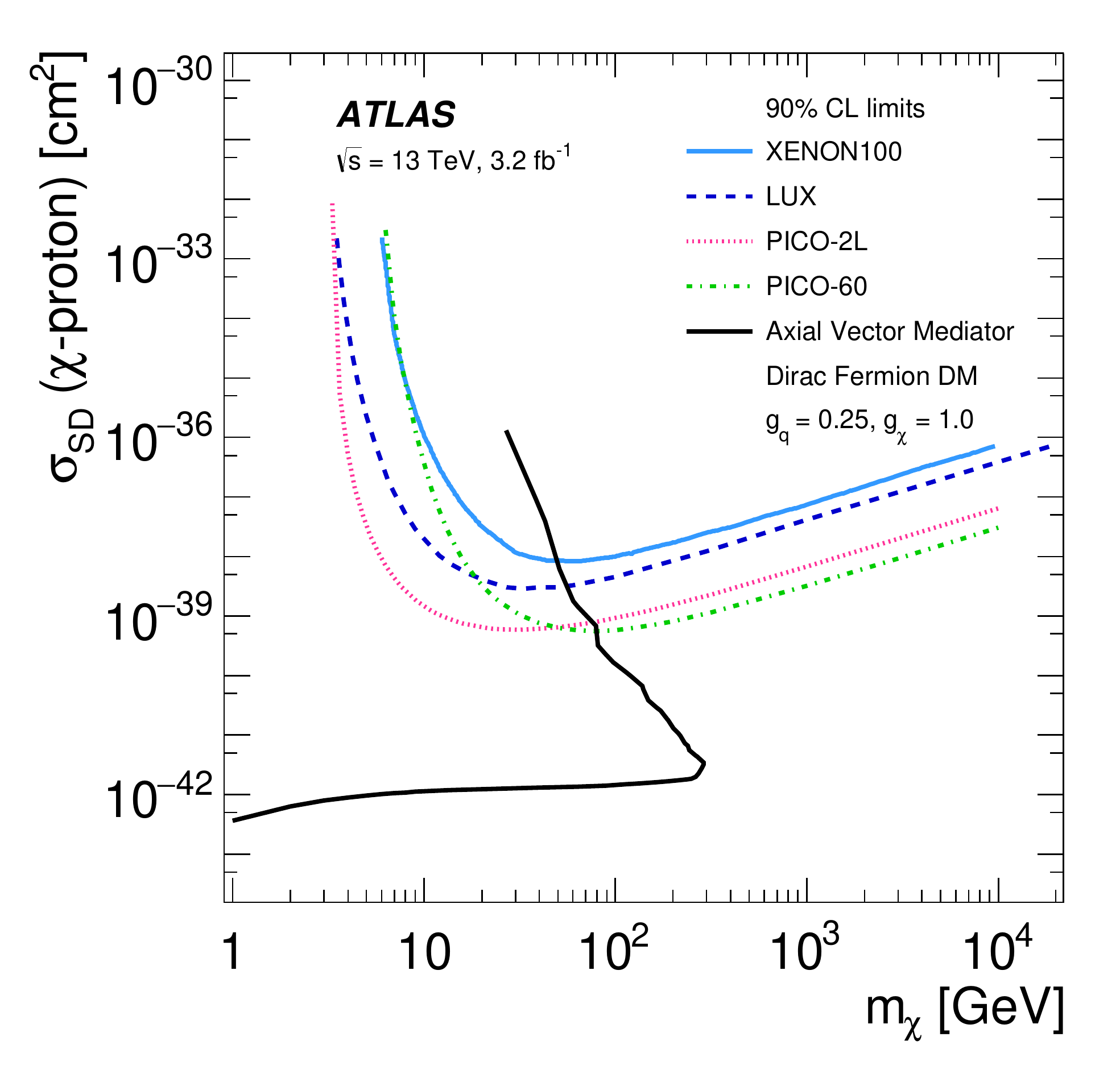}
\caption{
Left: 95\% CL exclusion contours in the $m_\chi$--$m_A$ parameter plane. The solid (dashed) curve shows the median of the observed (expected) limit, while the bands 
indicate the $\pm1\sigma$  theory uncertainties in the observed limit and $\pm1\sigma$ range of the expected limit in the absence of a signal. 
The red curve corresponds to the expected relic density.
The region excluded due to perturbativity, defined by $m_\chi > \sqrt{\pi/2}\ m_A$, is indicated by the hatched area.
Right:
A comparison of the inferred limits to the constraints from direct detection experiments on the spin-dependent WIMP--proton scattering cross section
in the context of the $Z'$-like simplified model with axial-vector couplings. Unlike in the $m_\chi$--$m_A$ parameter plane, 
the limits are shown at 90\% CL. The results from this analysis, excluding the region to the left of the contour, are compared with limits from the XENON100 \cite{Aprile:2013doa}, LUX \cite{Akerib:2016lao}, and PICO \cite{Amole:2015pla,Amole:2016pye} experiments.
The comparison is model-dependent and solely valid in the context of this model, assuming minimal mediator width and the coupling values $g_q=1/4$ and $g_\chi=1$.
}
\label{fig:dm}
\end{figure}

\FloatBarrier

\section{Conclusions}
\label{sec:sum}

In summary, results are reported from a search for new phenomena in 
events with an energetic jet and large missing transverse momentum 
in proton--proton collisions at $\sqrt{s}=13\TeV$ at the LHC,
based on data corresponding to an integrated 
luminosity of $3.2\,\ifb$ collected by the ATLAS experiment in 2015. 
The measurements are in agreement with the 
SM predictions.

The results are translated into 
model-independent 95$\%$ confidence-level upper limits
on $\sigma \times A \times \epsilon$ in the range 
 553--19~fb, depending on the selection criteria considered. The results are presented in terms
of lower limits on the fundamental Planck scale, $M_D$, versus the number of extra spatial dimensions 
in  the ADD LED model.
Values of $M_D$ below $6.58\TeV$ at $n=2$ and below $4.31\TeV$ at $n=6$ are excluded at
95$\%$~CL.   
Similarly, the results are interpreted in terms of the search for 
squark pair production in a compressed supersymmetric scenario. 
In the case of stop and sbottom pair production with $\tilde{t}_1 \to  c +  \tilde{\chi}^{0}_{1}$
and $\tilde{b}_1 \to  b +  \tilde{\chi}^{0}_{1}$, respectively,  squark masses 
below $323\GeV$ are excluded at 95$\%$~CL. In the case of squark pair production with
$\tilde{q} \to  q +  \tilde{\chi}^{0}_{1}  \ (q=u,d,c,s)$  squark masses below $608\GeV$ are excluded.  
Altogether, these results extend the exclusion from previous analyses at the LHC.

Finally, the results are interpreted in terms of upper limits on the  
pair-production cross section 
of WIMPs. A simplified model is used with an axial-vector mediator, given couplings to 
fermions $g_\chi=1$ and $g_q=1/4$,  and considering Dirac fermions as dark matter candidates.
Mediator masses below $1\TeV$ are excluded at 95$\%$~CL for WIMP masses below $250\GeV$.   
These results are translated, in a model-dependent manner, into upper limits on spin-dependent
contributions to the WIMP--nucleon elastic cross section as a function of the WIMP mass.
WIMP--proton cross sections above $10^{-42}$~cm${}^2$ are excluded at 90$\%$ CL for WIMP masses below $10\GeV$, complementing results from direct-detection experiments.

\section*{Acknowledgments}


We thank CERN for the very successful operation of the LHC, as well as the
support staff from our institutions without whom ATLAS could not be
operated efficiently.

We acknowledge the support of ANPCyT, Argentina; YerPhI, Armenia; ARC, Australia; BMWFW and FWF, Austria; ANAS, Azerbaijan; SSTC, Belarus; CNPq and FAPESP, Brazil; NSERC, NRC and CFI, Canada; CERN; CONICYT, Chile; CAS, MOST and NSFC, China; COLCIENCIAS, Colombia; MSMT CR, MPO CR and VSC CR, Czech Republic; DNRF and DNSRC, Denmark; IN2P3-CNRS, CEA-DSM/IRFU, France; GNSF, Georgia; BMBF, HGF, and MPG, Germany; GSRT, Greece; RGC, Hong Kong SAR, China; ISF, I-CORE and Benoziyo Center, Israel; INFN, Italy; MEXT and JSPS, Japan; CNRST, Morocco; FOM and NWO, Netherlands; RCN, Norway; MNiSW and NCN, Poland; FCT, Portugal; MNE/IFA, Romania; MES of Russia and NRC KI, Russian Federation; JINR; MESTD, Serbia; MSSR, Slovakia; ARRS and MIZ\v{S}, Slovenia; DST/NRF, South Africa; MINECO, Spain; SRC and Wallenberg Foundation, Sweden; SERI, SNSF and Cantons of Bern and Geneva, Switzerland; MOST, Taiwan; TAEK, Turkey; STFC, United Kingdom; DOE and NSF, United States of America. In addition, individual groups and members have received support from BCKDF, the Canada Council, CANARIE, CRC, Compute Canada, FQRNT, and the Ontario Innovation Trust, Canada; EPLANET, ERC, FP7, Horizon 2020 and Marie Sk{\l}odowska-Curie Actions, European Union; Investissements d'Avenir Labex and Idex, ANR, R{\'e}gion Auvergne and Fondation Partager le Savoir, France; DFG and AvH Foundation, Germany; Herakleitos, Thales and Aristeia programmes co-financed by EU-ESF and the Greek NSRF; BSF, GIF and Minerva, Israel; BRF, Norway; Generalitat de Catalunya, Generalitat Valenciana, Spain; the Royal Society and Leverhulme Trust, United Kingdom.

The crucial computing support from all WLCG partners is acknowledged
gratefully, in particular from CERN and the ATLAS Tier-1 facilities at
TRIUMF (Canada), NDGF (Denmark, Norway, Sweden), CC-IN2P3 (France),
KIT/GridKA (Germany), INFN-CNAF (Italy), NL-T1 (Netherlands), PIC (Spain),
ASGC (Taiwan), RAL (UK) and BNL (USA) and in the Tier-2 facilities
worldwide.


\printbibliography


 \newpage 
\begin{flushleft}
{\Large The ATLAS Collaboration}

\bigskip

M.~Aaboud$^\textrm{\scriptsize 136d}$,
G.~Aad$^\textrm{\scriptsize 87}$,
B.~Abbott$^\textrm{\scriptsize 114}$,
J.~Abdallah$^\textrm{\scriptsize 65}$,
O.~Abdinov$^\textrm{\scriptsize 12}$,
B.~Abeloos$^\textrm{\scriptsize 118}$,
R.~Aben$^\textrm{\scriptsize 108}$,
O.S.~AbouZeid$^\textrm{\scriptsize 138}$,
N.L.~Abraham$^\textrm{\scriptsize 150}$,
H.~Abramowicz$^\textrm{\scriptsize 154}$
% ...
\bigskip
\\
% ...
$^{12}$ Institute of Physics, Azerbaijan Academy of Sciences, Baku, Azerbaijan\\
% ...
$^{65}$ University of Iowa, Iowa City IA, United States of America\\
% ...
$^{87}$ CPPM, Aix-Marseille Universit{\'e} and CNRS/IN2P3, Marseille, France\\
% ...
$^{108}$ Nikhef National Institute for Subatomic Physics and University of Amsterdam, Amsterdam, Netherlands\\
% ...
$^{114}$ Homer L. Dodge Department of Physics and Astronomy, University of Oklahoma, Norman OK, United States of America\\
% ...
$^{118}$ LAL, Univ. Paris-Sud, CNRS/IN2P3, Universit{\'e} Paris-Saclay, Orsay, France\\
% ...
$^{136}$ $^{(a)}$ Facult{\'e} des Sciences Ain Chock, R{\'e}seau Universitaire de Physique des Hautes Energies - Universit{\'e} Hassan II, Casablanca; $^{(b)}$ Centre National de l'Energie des Sciences Techniques Nucleaires, Rabat; $^{(c)}$ Facult{\'e} des Sciences Semlalia, Universit{\'e} Cadi Ayyad, LPHEA-Marrakech; $^{(d)}$ Facult{\'e} des Sciences, Universit{\'e} Mohamed Premier and LPTPM, Oujda; $^{(e)}$ Facult{\'e} des sciences, Universit{\'e} Mohammed V, Rabat, Morocco\\
% ...
$^{138}$ Santa Cruz Institute for Particle Physics, University of California Santa Cruz, Santa Cruz CA, United States of America\\
% ...
$^{150}$ Department of Physics and Astronomy, University of Sussex, Brighton, United Kingdom\\
% ...
$^{154}$ Raymond and Beverly Sackler School of Physics and Astronomy, Tel Aviv University, Tel Aviv, Israel\\
% ...
$^{a}$ Also at Department of Physics, King's College London, London, United Kingdom\\
$^{b}$ Also at Institute of Physics, Azerbaijan Academy of Sciences, Baku, Azerbaijan\\
$^{c}$ Also at Novosibirsk State University, Novosibirsk, Russia\\
$^{d}$ Also at TRIUMF, Vancouver BC, Canada\\
$^{e}$ Also at Department of Physics {\&} Astronomy, University of Louisville, Louisville, KY, United States of America\\
% ...
\end{flushleft}



\end{document}